\documentclass[preprint,superscriptaddress,nofootinbib,preprintnumbers,amsmath,amssymb]{revtex4-1}

\usepackage{hyperref}
\usepackage{amsfonts}
\usepackage{mathrsfs}
\usepackage{epsfig}
\usepackage{graphicx}               
\usepackage{url}
\usepackage{hyperref}
\usepackage{float}
\usepackage{color}

\usepackage{graphicx}
\usepackage{epsf}

\usepackage{comment}

\newcommand{\nc}{\newcommand}

\nc{\beq}{\begin{equation}}
\nc{\eeq}{\end{equation}}
\nc{\beqa}{\begin{eqnarray}}
\nc{\eeqa}{\end{eqnarray}}

\usepackage{slashed}

\newcommand{\lsim}{\!\mathrel{\hbox{\rlap{\lower.55ex \hbox{$\sim$}} \kern-.34em \raise.4ex \hbox{$<$}}}}
\newcommand{\gsim}{\!\mathrel{\hbox{\rlap{\lower.55ex \hbox{$\sim$}} \kern-.34em \raise.4ex \hbox{$>$}}}}

\newcommand{\vev}[1]{ \left\langle {#1} \right\rangle }
\newcommand{\abs}[1]{ \left| {#1} \right| }
\def\be{\begin{equation}}
\def\ee{\end{equation}}
\def\ie{{\it i.e.}}
\def\eg{{\it e.g.}}
\def\GeV{\text{ GeV}}
    
\def\Lmax{\Lambda_{\rm max}}
\DeclareMathOperator{\erf}{erf}
\newcommand{\Sref}[1]{Sec.~\ref{#1}}
\newcommand{\Fref}[1]{Fig.~\ref{#1}}
\newcommand{\Eref}[1]{Eq.~(\ref{#1})}
\newcommand{\Rref}[1]{Ref.~\cite{#1}}

\newcommand\affspc{\vspace{4pt}}

\def\hcld{h_{\rm cl}}
\def\hsrb{h_{\slashed{sr}}} 

\renewcommand{\(}{\left(}
\renewcommand{\)}{\right)}

\usepackage{footmisc}
\usepackage{setspace}

\begin{document}

\preprint{FERMILAB-PUB-16-250-T}
\preprint{MCTP-16-15}

\title{Spacetime Dynamics of a Higgs Vacuum Instability During Inflation}

\author{William E.\ East}
\affiliation{Kavli Institute for Particle Astrophysics and Cosmology, Stanford
University, SLAC National Accelerator Laboratory, Menlo Park, CA 94025, USA \affspc}
\affiliation{Perimeter Institute for Theoretical Physics, Waterloo, ON N2L 2Y5, Canada \affspc}
\author{John Kearney}
\affiliation{Theoretical Physics Department, Fermi National Accelerator Laboratory, Batavia, IL 60510, USA \affspc}
\author{Bibhushan Shakya}
\affiliation{Michigan Center for Theoretical Physics, University of Michigan, Ann Arbor, MI 48019, USA \affspc}
\author{Hojin Yoo}
\affiliation{Theory Group, Lawrence Berkeley National Laboratory and Berkeley Center for Theoretical Physics, University of California, Berkeley, CA 94709, USA \affspc}
\author{Kathryn M. Zurek}
\affiliation{Theory Group, Lawrence Berkeley National Laboratory and Berkeley Center for Theoretical Physics, University of California, Berkeley, CA 94709, USA \affspc}

\begin{abstract}

A remarkable prediction of the Standard Model is that, in the absence of
corrections lifting the energy density, the Higgs potential becomes negative at
large field values.  
If the Higgs field samples this part of the potential during inflation, the negative energy density may locally destabilize the spacetime.
We use numerical simulations of the Einstein equations to study the evolution of inflation-induced Higgs fluctuations as they grow towards the true (negative-energy) minimum.  These simulations show that forming a single patch of true vacuum in our past light cone during inflation is incompatible with the existence of our Universe; the boundary of the true vacuum region grows outward in a causally disconnected manner from the crunching interior, which forms a black hole.  
We also find that these black hole horizons may be arbitrarily elongated---even forming black strings---in violation of the hoop conjecture.
By extending the numerical solution of the Fokker-Planck equation to the exponentially suppressed tails of the field distribution at large field values, we derive a rigorous correlation between a future measurement of the tensor-to-scalar ratio and the scale at which the Higgs potential must receive stabilizing corrections in order for the Universe to have survived inflation until today.   

\end{abstract}

\maketitle

\section{Introduction}

A striking feature of the Standard Model (SM) is that, in the absence of
stabilizing corrections, the Higgs potential develops an instability, with the maximum of the potential occurring at
$V(\Lmax)^{1/4} \sim 10^{10} \GeV$.
This leads to the existence of a ``true vacuum'' at large Higgs field values,
which may carry important consequences for our Universe
\cite{Sher:1988mj,Sher:1993mf,Casas:1994qy,Altarelli:1994rb,Ellis:2009tp,EliasMiro:2011aa,Isidori:2007vm,Isidori:2001bm,Bezrukov:2012sa,Degrassi:2012ry,Buttazzo:2013uya}. 
Our present existence does not necessarily demand physics beyond the
SM, since current measurements of the Higgs boson and top quark masses indicate that
the electroweak (EW) vacuum is metastable, \ie, long-lived relative to the age
of the Universe.  The scenario is different, however, if our Universe underwent
an early period of cosmic inflation with substantial energy density.  The inflaton
energy density, parametrized by the Hubble parameter $H$, produces
large local fluctuations in the Higgs field, $\delta h \sim \frac{H}{2 \pi}$.
As such, when $H$ is sufficiently large during inflation, the Higgs field may sample the unstable part of the potential.

If sampling this part of the potential can be shown to be catastrophic for the
surrounding spacetime, the
eventual survival of our Universe in the EW vacuum would consequently imply constraints on the
nature of the inflationary epoch that gave rise to our Universe.  Conversely,
near-future cosmic microwave background experiments will probe
tensor-to-scalar ratios of $r \gsim 0.002$ \cite{Creminelli:2015oda}, corresponding to inflationary scales
$H > 10^{13} \GeV$. If it can be shown that the
SM Higgs potential is inconsistent with such high-scale inflation, a measurement of nonzero $r$ provides
evidence for the existence of stabilizing corrections to the Higgs potential. 

In recent years, the interplay between the SM Higgs potential instability and inflation has
received significant attention 
\cite{Espinosa:2007qp,Lebedev:2012sy,Kobakhidze:2013tn,Enqvist:2013kaa,Hook:2014uia,Enqvist:2014bua,Herranen:2014cua,Kobakhidze:2014xda,Fairbairn:2014zia,Shkerin:2015exa,Kearney:2015vba,Espinosa:2015qea}.
A complete treatment of this problem has two important aspects: first, the evolution of
the Higgs field under a combination of (inflation-induced) quantum fluctuations
and the classical potential and, second, the evolution of spacetime responding to the
Higgs vacuum. 

Initial groundwork on the first aspect was laid in \Rref{Espinosa:2007qp}, which
employed a stochastic, Fokker-Planck (FP) approach to study the evolution and
distribution of Higgs fluctuations in Hubble-sized patches during inflation.
While this is a suitable approach incorporating both leading classical and
quantum effects, the analysis of \cite{Espinosa:2007qp} was predicated on the
assumption that fluctuations exceeding $\Lmax$ rapidly transitioned to the true
vacuum and disappeared, resulting in a miscalculation of the distribution.  It
was subsequently shown in \cite{Hook:2014uia}, however, that fluctuations
continue to evolve in an inflationary background well past the point where the
Higgs quartic becomes negative. In fact, it is the formation of fluctuations
well beyond $\Lmax$ that carry the most significant implications for our
Universe, making it necessary to study the full distribution of fluctuations. As \Rref{Kearney:2015vba} later demonstrated, a true vacuum patch capable of
backreacting on the inflating spacetime only forms at about the time that a
fluctuation becomes sufficiently large that the Higgs field locally exits the
slow-roll regime.

The first meaningful investigation of the second aspect---the response of the spacetime to the Higgs vacuum evolution---appeared in \Rref{Espinosa:2015qea}.\footnote{Earlier studies did not investigate the reaction of the spacetime, instead assuming a variety of outcomes. For example, \cite{Espinosa:2007qp} assumed that fluctuations to the true vacuum only locally terminate inflation, rapidly forming AdS regions that ``benignly" crunch (shrinking to negligible volume), while \cite{Kobakhidze:2013tn,Fairbairn:2014zia} supposed a single true vacuum patch in our past light cone eventually devours all of spacetime. Reference~\cite{Hook:2014uia} considered both extreme possibilities.}
In order to make the study analytically tractable, they adopted an idealized setup of a spherically symmetric thin-wall anti-de Sitter (AdS) bubble in a de Sitter (dS) background and found that true vacuum bubbles persist throughout inflation for realistic parameters. As such, the formation of a single such true vacuum patch in our past light cone during inflation would be disastrous for our Universe---after inflation, such patches would expand and destroy the surrounding space in the EW vacuum.\footnote{See also \cite{Freivogel:2007fx,Johnson:2010bn} for related working on the collision of crunching bubbles.}

The main goal of this paper is a comprehensive study of both aspects, the field evolution and subsequent reaction of the spacetime.
We improve the study of the former aspect by numerically resolving the full probability distribution of Higgs fluctuations in the FP equation, even into the exponentially suppressed tails that govern single patches in our past light cone.
This is in contrast to previous studies \cite{Espinosa:2007qp,Hook:2014uia,Espinosa:2015qea}, which relied on a type of ``matching'' procedure between quantum-dominated and classical-dominated evolution in the FP treatment.\,\footnote{This matching procedure consists of using the FP equation to track the field evolution to the point where classical effects start to dominate over quantum effects, and switching to the classical equation of motion beyond this point (thus ignoring the quantum effects) to track the subsequent evolution.}
We carry out a comprehensive study of the second aspect by employing full numerical solutions to the Einstein equations instead of the thin-wall approximation.

Moving beyond the approximations previously
employed in the literature is vital to providing a complete description of the
interplay between inflation and the Higgs field for several reasons. First, a more complete numerical solution to
the FP equation allows us to fully capture the important effects of the
renormalization group-improved potential, as well as the crucial non-Gaussian
tails of the Higgs field value distribution. In particular, as
\Rref{Kearney:2015vba} argued based on the Higgs effective potential in dS space
calculated in \cite{Herranen:2014cua} and Wilsonian effective field theory, an appropriate scale at which to evaluate the Higgs self-coupling is $\mu \simeq \sqrt{H^2 + h^2}$ as opposed to $\mu
\simeq \abs{h}$. This choice minimizes large logarithms and incorporates
the relevant energy scale from inflation.  As we shall see, fully including the
effects of the renormalization group-improved potential influences both small
and large fluctuations.
Meanwhile, as we demonstrate, it is the exponentially suppressed but long tails of the distribution that ultimately determine the rate at which true vacuum patches form.

Second, since the evolution of a Higgs fluctuation becomes classical well before becoming sufficiently large to backreact on the spacetime, it is important to study a patch rapidly evolving to the true vacuum, gaining significant energy as it falls, as a dynamical general relativity process.
The thin-wall approximation employed in \cite{Espinosa:2015qea} is valid when fluctuations beyond the potential barrier at $\Lmax$ occur via a Coleman-de Luccia tunneling process \cite{Coleman:1980aw}, resulting in a true vacuum bubble interior that rapidly transitions to false vacuum exterior across a thin boundary.
During inflation, however, Higgs fluctuations are more appropriately described by a
broad, Hubble-sized variation in the field, more akin to a Hawking-Moss 
instanton \cite{Hawking:1981fz} (see \cite{Hook:2014uia} for a detailed
discussion).  
Here we will not make any simplifying assumptions regarding the Higgs
fluctuation being a region of AdS separated from the surround dS at an
infinitely thin bubble wall, though we will still use the term ``bubble" to
refer to dynamically formed regions where the Higgs field is near the true vacuum. 
Our numerical simulations allow us to study the full behavior of extended fluctuations, offering the first in-depth understanding of the field and spacetime dynamics of these Higgs fluctuations. 

In particular, we highlight three important aspects of true vacuum patch formation. 
First, we show that patches only rapidly diverge to the true vacuum and backreact on the inflating spacetime once the Higgs field locally exits the slow-roll regime.
Second, the associated large negative energy density does terminate inflation locally, eventually producing a crunching region, but this region is hidden behind a black hole horizon that is surrounded by an expanding shell of negative energy density.
Third, for reasonable parameters, the shell of negative energy density expands into the surrounding spacetime in a manner causally disconnected from the crunching interior, in contrast to the thin-wall AdS bubble. As a result, its growth is not sensitive to the crunching behavior of the spacetime in the interior, allowing such true vacuum regions to persist through inflation.

We thus confirm that the formation of a single, sufficiently large fluctuation during inflation precludes the existence of our Universe, resulting in a bound $H/\Lmax \lsim 0.07$ that, once a number of competing effects are taken into account, is similar to that found in previous studies~\cite{Hook:2014uia,Espinosa:2015qea}.  
In addition, our numerical approach enables us to study more complicated nonspherical solutions, where we find that the formation of AdS-like regions from the field falling to the true minimum at negative potential energy allows for the formation of black holes with arbitrarily elongated horizons (and black strings), in violation of the hoop conjecture~\cite{thorne_hoop}.  

We emphasize that, while the presence of new physics at the weak scale
could substantially change the quantitative features of the Higgs evolution due to the
modified Higgs potential, there are many conceptual points in the interplay
between an inflating spacetime and a field with a vacuum instability that are applicable in a wider context.  Furthermore, we illustrate in this
work some of the qualitatively different features that Einstein gravity exhibits
in the presence of negative energy density, including the formation of black
holes with arbitrarily elongated horizons, or even black strings, that hide the
crunching regions from outside observers.  These touch on fundamental
considerations in gravity such as the topology of black hole horizons, the hoop
conjecture, and cosmic censorship.

The rest of this paper is organized as follows.  In \Sref{evolution_stages}, we briefly review the stochastic approach to studying the evolution of Higgs field fluctuations using the FP equation.
\Sref{numerical_approach} is the main part of this paper where, using full numerical simulations, we study the spacetime dynamics of the patches exhibiting large fluctuations that evolve to the true vacuum.
In \Sref{fp_limits}, we present a complete numerical solution of the FP equation, allowing us to extract constraints on the Hubble scale or the form of the Higgs potential from the survival of our Universe through inflation. 
Finally, we summarize our conclusions in \Sref{conclusions}. 

\section{Evolution of the Higgs Field During Inflation}
\label{evolution_stages}

To set the stage for studying the evolution of spacetime in response to unstable
Higgs fluctuations in the next section, here we review the evolution of the Higgs field during inflation and the formation of large fluctuations as modeled by the Fokker-Planck (FP) equation. We restrict ourselves here to providing the context;  a more quantitative numerical solution of the FP equation and in-depth discussion will be presented later in \Sref{fp_limits}.

When $H^2 \gsim V''(\Lambda_{\rm max})$, where $V(\Lambda_{\rm max})$ is the
maximum of the potential, a statistical approach can be utilized for studying
the Higgs evolution during inflation via the Fokker-Planck (FP) equation
\cite{Linde:1991sk, starobinsky:stochastic,Espinosa:2007qp}, 
\be
\label{Eq:FP}
\frac{\partial P}{\partial t} = \frac{\partial}{\partial h}\left[\frac{V'(h)}{3H} P + \frac{H^3}{8 \pi^2}\frac{\partial P}{\partial h} \right].
\ee
This equation governs the probability distribution $P(h,t)$ corresponding to the
average field value $h$ in a patch of size $\sim H^{-1}$ at time $t$. The second moment of the distribution, $\langle h^2 \rangle = \int dh h^2 P(h,t)$, reproduces well the behavior obtained from the equation of motion for $\langle h^2 \rangle$ in the Gaussian approximation with subhorizon modes integrated out, at least for small fluctuations (see, \eg, \cite{Finelli:2008zg,Kearney:2015vba}).\footnote{For a discussion of the regime of validity of this equation for inflationary evolution of the Higgs (as opposed to a Coleman--de Lucia or Hawking-Moss instanton solution), we refer the interested reader to \Rref{Hook:2014uia}.}

The first term on the right of \Eref{Eq:FP} accounts for classical evolution due to the
potential in the slow-roll approximation.  As argued in \Rref{Kearney:2015vba},
since the FP equation describes the evolution of Higgs fluctuations on scales $R
\gsim H^{-1}$, the potential $V$ appearing in \Eref{Eq:FP} is an effective
potential containing only superhorizon modes.  Mode functions of non-Higgs SM
fields (fermions and gauge bosons) rapidly decay outside the horizon, so these
fields do not correct this infrared/superhorizon Higgs effective potential. However,
they do renormalize the quartic coupling in the ultraviolet (UV).  As such, the appropriate
potential is
\be
\label{eq:VhforFP}
V(h) = \frac{\lambda (\mu)}{4} h^4
\ee
where $h$ is the canonically normalized Higgs field and at leading order $\lambda(\mu)$ is the RG-improved quartic, matched to the UV quartic (taken to be the SM quartic as
in Minkowski space) at the scale at which the SM fields decouple. Taking the
Higgs-dependent mass into account, the optimal choice of scale is $\mu \simeq
\sqrt{H^2 + h^2}$ \cite{Herranen:2014cua,Kearney:2015vba}---for small fluctuations,
this corresponds to the infrared cutoff $\mu \simeq H$ below which subhorizon physics is
integrated out \cite{Starobinsky:1994bd}.
We assume for the time being that the Higgs has no corrections to
\Eref{eq:VhforFP} from, \eg, a coupling to gravity of the form $H^2 h^2$; we
return to the impact of such a term in \Sref{subsec:stabilizingcorrections}.

The second term in \Eref{Eq:FP} corresponds to the quantum-to-classical
transition experienced by field modes during inflation as a result of horizon
crossing. The result is a random walk for $h$ with steps of order $\sim
\frac{H}{2 \pi}$ as subsequent modes cross the horizon.  These steps can
also be thought of as thermal fluctuations in an inflationary background with a
Gibbons-Hawking temperature $\frac{H}{2\pi}$ \cite{Brown:2007sd}, which increase
or decrease the size of a local fluctuation depending on whether the modes
crossing the horizon add coherently or destructively with the longer wavelength
modes that froze out earlier. Thus the characteristic size of the spatial field
structure induced by such fluctuations is $\; \sim H^{-1}$.

Initially, evolution is dominated by quantum fluctuations via the second term.
This causes large local fluctuations in the Higgs field that, for sufficiently
large $H$, may result in the field locally sampling the unstable part of the
potential, $\abs{h} \gsim \Lmax$.  Though a positive quartic may somewhat
suppress the growth of fluctuations for $\sqrt{H^2 + h^2} \lsim \Lmax$, the classical effect due to the unstable
potential causes the distribution to grow somewhat more quickly once
$\sqrt{H^2 + h^2} \gsim \Lmax$.  At this point, the stochastic term still dominates over the classical potential, so that a fluctuation does not yet grow inexorably toward the true vacuum.  This only happens once the classical potential comes to dominate over quantum effects, when $\abs{h} \gsim \hcld$,
where 
\begin{align}
\label{eq:cl}
V'(\hcld) & = -\frac{3 H^3}{2 \pi}, & \hcld & \approx H\left( \frac{3}{-2\pi \lambda}\right)^{1/3},
\end{align}
\ie, when the slow-roll evolution due to the potential, $\dot{h}\Delta t\simeq V'(h_{\rm cl})/(3 H^2)$, exceeds the stochastic evolution due to inflationary fluctuations. 
From this point, as described in \cite{Hook:2014uia}, the field necessarily diverges to the true vacuum.  

However, as first emphasized in \cite{Kearney:2015vba}, the local energy density
at this point is still overwhelmingly dominated by the inflationary energy
density. Due to Hubble friction, the field continues to undergo slow-roll
evolution, and a significant number of $e$-folds must pass after entering the
classical regime, \Eref{eq:cl}, until the Higgs exits the slow-roll regime.
Meanwhile, inflation proceeds unabated.  This regime is therefore still well
described by the Fokker-Planck equation.  It is only after a fluctuation becomes
very large, $\abs{h} \gsim \hsrb$, where 
\be
\hsrb = -\frac{V'(\hsrb)}{3 H^2} \approx H\left( \frac{3}{-\lambda} \right)^{1/2},
\label{eqn_hsr}
\ee
that the slow-roll approximation breaks down, and the fluctuation rapidly
diverges to the true vacuum.  Only then does the energy density in the Higgs field
become sufficiently large to backreact on the spacetime---we will explore this
backreaction in the subsequent section.  Consequently, to determine the fraction of
patches that reach the true vacuum and backreact during inflation, one must
track the evolution of the fluctuations to $\abs{h} \sim \hsrb$.

Before meaningful conclusions can be drawn from the solution of the FP equation,
we must first understand how true vacuum patches form and evolve in spacetime as
inflation proceeds and eventually ends. In the next section, we investigate this
question with full dynamical simulations, before returning to the solution of
the FP equation and discussing the implications for inflation in the
subsequent section. 

\section{Higgs and Spacetime Dynamics}
\label{numerical_approach}

This section, which comprises the main part of the paper, presents the results of
general relativistic simulations
to study the classical spacetime and field dynamics
of Higgs fluctuations during inflation. In \Sref{subsec:setup} we outline
our setup and methods for solving the field equations.  We evaluate the
time scale for a Higgs fluctuation to fall to the true vacuum in
\Sref{subsec:falling}, and illustrate that the spatial extent of the field
has a negligible effect on this in the relevant parts of parameter space.
In \Sref{subsec:true_vacuum} we study the formation of a region of true vacuum. We find that the crunching region is hidden behind a black hole horizon,
which is itself surrounded by an expanding region of negative energy density. In
\Sref{subsec:bubble}, we examine the growth of the regions rapidly evolving to the true Higgs vacuum, demonstrating that it is generically causally disconnected from the the noninflating interior.  As such, once formed, these regions expand and persist throughout inflation.
\Sref{subsec:nonspherical} is devoted to an examination of nonspherically symmetric Higgs fluctuations, illustrating how they can form arbitrarily elongated black holes and black strings by virtue of the negative potential energy of the Higgs field.

\subsection{Numerical setup}
\label{subsec:setup}

To model the classical evolution of the Higgs field, we consider a scalar field
$h$, with equation of motion $\Box h = V'$ (where, in terms of covariant derivatives, $\Box \equiv \nabla_a \nabla^a$), coupled to the Einstein field
equations. For the purposes of the simulation, we add a Planck-suppressed operator to \Eref{eq:VhforFP} to stabilize the potential at large field values,
\beq
V(h) = \Lambda_{\rm Infl}+\frac{\lambda}{4} h^4 + \frac{\lambda_6}{6M_P^2}h^6
. 
\eeq
Here the constant term represents the contribution from the inflaton and is
related to the Hubble scale during inflation as $\Lambda_{\rm Infl}=3M_P^2H^2$,
where $M_P$ is the reduced Planck mass.  The parameters $\lambda<0$ and
$\lambda_6>0$ are constants representing, respectively, the effective
quartic term in the instability regime and some unknown higher-dimensional
correction. The higher-dimensional correction generates a Planck-scale global minimum at
\beq
\frac{h_{\rm min}}{M_{P}}=0.1\(\frac{\lambda}{-0.01}\)^{1/2}
\left(\lambda_6 \right)^{-1/2}
\eeq
with value 
\beq
\frac{V_{\rm min}}{\Lambda_{\rm
Infl}}=1-1.85\times10^5\(\frac{\lambda}{-0.01}\) \(\frac{\Lambda_{\rm Infl}}{(10^{16}\ {\rm
GeV})^4}\)^{-1} \(\frac{h_{\rm min}}{0.1 M_P}\)^4.
\eeq
For $h_{\rm min} \sim M_P$, $-V_{\rm min}\gg \Lambda_{\rm Infl}$. However, depending on the energy scale of inflation, the magnitude of the higher-order coupling, and the exact value of $\lambda$ within the experimental error
bars, it is conceivable that $h_{\rm min}\ll M_P$ or $h_{\rm min} \sim M_P$, as
well as that $-V_{\rm min}\gg \Lambda_{\rm Infl}$ or $-V_{\rm min}\sim
\Lambda_{\rm Infl}$. 
Keeping to cases with a negative energy density true vacuum with $V_{\rm min} \leq -\Lambda_{\rm Infl}$, 
we have considered all hierarchies and found that our main results do not depend strongly on the values of these parameters.\footnote{Note that the case where the true vacuum energy density does not exceed the inflationary energy density assumedly constitutes a worst-case scenario in which regions that transition to the true vacuum continue to inflate. So, such regions certainly persist throughout inflation, allowing them to nucleate and destroy any remaining space in the EW vacuum afterwards.}
Hence, for presenting our results, we mainly choose the values of these
parameters based on computational expediency without worrying about covering the
entire physically viable parameter space; below we will use the default parameters $h_{\rm min}=0.1M_P$
and $V_{\rm min}/\Lambda_{\rm Infl}=-100$ unless otherwise stated.

For initial conditions, we consider Higgs fluctuations 
that are momentarily static, $\partial_t{h}=0$, and have
axisymmetric spatial profiles. 
We consider both Gaussian spatial profiles given by
\begin{align}
h(x,y,z) & = h_{\rm in} e^{-\rho^2/2}, & \rho^2 & = \frac{(x^2+y^2)}{R^2_{xy}} + \frac{z^2}{R_z^2}, 
\label{eqn:gauss}
\end{align}
as well as, for illustrative purposes, compactly supported, step-function-like profiles given by
\beq
h=
\left\{
	\begin{array}{ll}
		h_{\rm in}  & \mbox{if } \rho < 0.9 \\
		0 & \mbox{if } \rho \geq 1
	\end{array}
\right.
\label{eqn:step}
\eeq
where the function smoothly interpolates in the range $0.9 \leq \rho < 1.0$.  We mainly concentrate on the spherically symmetric case with $R_{xy}=R_z\equiv R$, but address nonspherical cases in \Sref{subsec:nonspherical}.
In presenting our results, we will make use of the coordinate radius $r_p=\sqrt{x^2+y^2+z^2}$, which will
closely match standard planar dS coordinates for regions where there has not been a strong backreaction on 
the spacetime---\ie, a factor of $e^{H t}$ should be applied to obtain the proper radius.

During the evolution, we search for and, in many cases, find marginally outer
trapped surfaces---apparent horizons---which signal the presence of black holes. In such cases, we excise the causally disconnected interiors of these surfaces from the numerical domain in order to continue the evolution outside the black holes. 
Further details of the implementation of the Einstein equations in our numerical simulations are described in the Appendix.

\subsection{Time scale to fall to the true vacuum}
\label{subsec:falling}

First, we examine the time scale for the field to fall into the true vacuum. Given an initial unstable fluctuation in the Higgs field, both gradient spreading and Hubble friction counter this process.  Taking a spherically symmetric
Gaussian profile with radius $R$, the value where the spatial Laplacian term and the gradient of the potential
term are equal and opposite at the maximum of the fluctuation is 
\beq
h_c = -V'(h_c)\frac{R^2}{3}.
\label{eqn_hc}
\eeq
Hence fluctuations with $h_{\rm in} \gsim h_c$ for a given $R$ will directly 
fall to the true vacuum, while those with $h_{\rm in} \lsim h_c$ 
will only fall to the true vacuum after the exponential expansion increases
their characteristic size, diluting the effects of spatial gradient terms.  

\begin{figure}
\begin{center}
\includegraphics[width=0.65\columnwidth,draft=false]{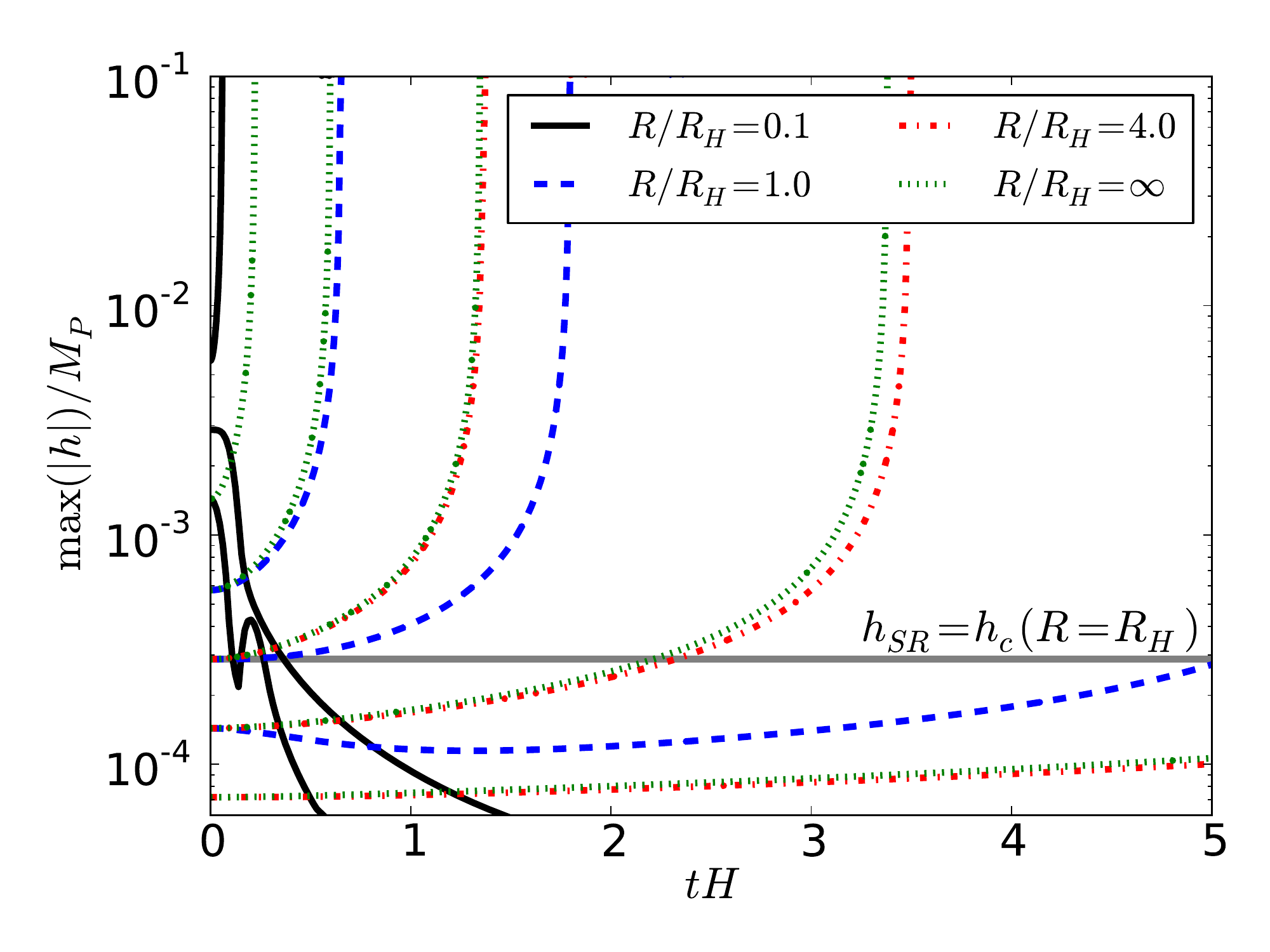}
\end{center}
\caption{Evolution of the maximum Higgs field value for 
initially spatially Gaussian fluctuations with size
$R/R_H=0.1$, 1.0, and 4.0 and various initial magnitudes. 
For comparison we also show the evolution of the amplitude when the 
initial radius is infinite (\ie, in an Friedmann-Roberston-Walker cosmology, with $R/R_H=\infty$; green dotted lines).
The horizontal grey line indicates the approximate value where the field exits
the slow-roll regime, which is also the approximate magnitude below which
gradient spreading is important for $R=R_H$. The features in the lowest black 
curve are due to the shifting location of the maximum field value in this case.
\label{fig:hmax} }
\end{figure}

Estimating the value below which Hubble friction prevents the field from falling
to the true vacuum within roughly a Hubble time (\ie, for the field to exit the
slow-roll regime) gives a similar value of $h_c$ as in \Eref{eqn_hc}, but
with $R=R_H\equiv H^{-1}$ [\Eref{eqn_hsr}].
Consequently, we expect fluctuations that have grown beyond $h_c$ and $\hsrb$ to fall swiftly
to the true vacuum. For smaller values, the evolution to the true vacuum takes several Hubble times or more,
during which the fluctuation becomes exponentially larger in spatial extent due to
inflation.

These expectations are corroborated by the simulations and illustrated in \Fref{fig:hmax}, where we plot the evolution of the maximum value of $\abs{h(r)}$ for initial Higgs fluctuations of various sizes and magnitudes.
Initial magnitudes larger than $\hsrb$ are found to fall rapidly into the true
vacuum within $\sim 1$ $e$-fold, and this behavior is largely insensitive to the
exact initial size of the fluctuation---in the plot, the curves of $R/R_H=1.0,
4.0$, and $\infty$ are very close together.

For spatially smaller fluctuations, the gradient terms and Hubble friction can
significantly slow down this process (lowest blue, red, and green lines) and, in
the extreme scenario of $R/R_H=0.1$, can even be dominant enough to force the field
back towards the EW vacuum even for $h > \hsrb$ (two lower black curves). 
However, the large Higgs fluctuations generated by inflation that are in danger of falling to the true vacuum are generated from superhorizon modes with $R > R_H$, which continue to inflate as they classically evolve to the true vacuum, resulting in a characteristic size $R \gg R_H$.
Hence we can safely assume that spatial variations have a negligible effect on the time for the development of a region of true vacuum for all realistic scenarios of interest.

\subsection{Regions of true vacuum and formation of black holes}
\label{subsec:true_vacuum}

Regions where the Higgs field fluctuations fall towards the true vacuum
experience a strong backreaction of the Higgs field on the spacetime. This
terminates inflation locally in these regions and, as the energy density becomes
negative, exponential expansion turns to contraction with the formation of a
crunching region. Here we examine the details of this process.

\begin{figure*}
\begin{center}
\hskip-0.5in
\includegraphics[width=0.55\columnwidth,draft=false]{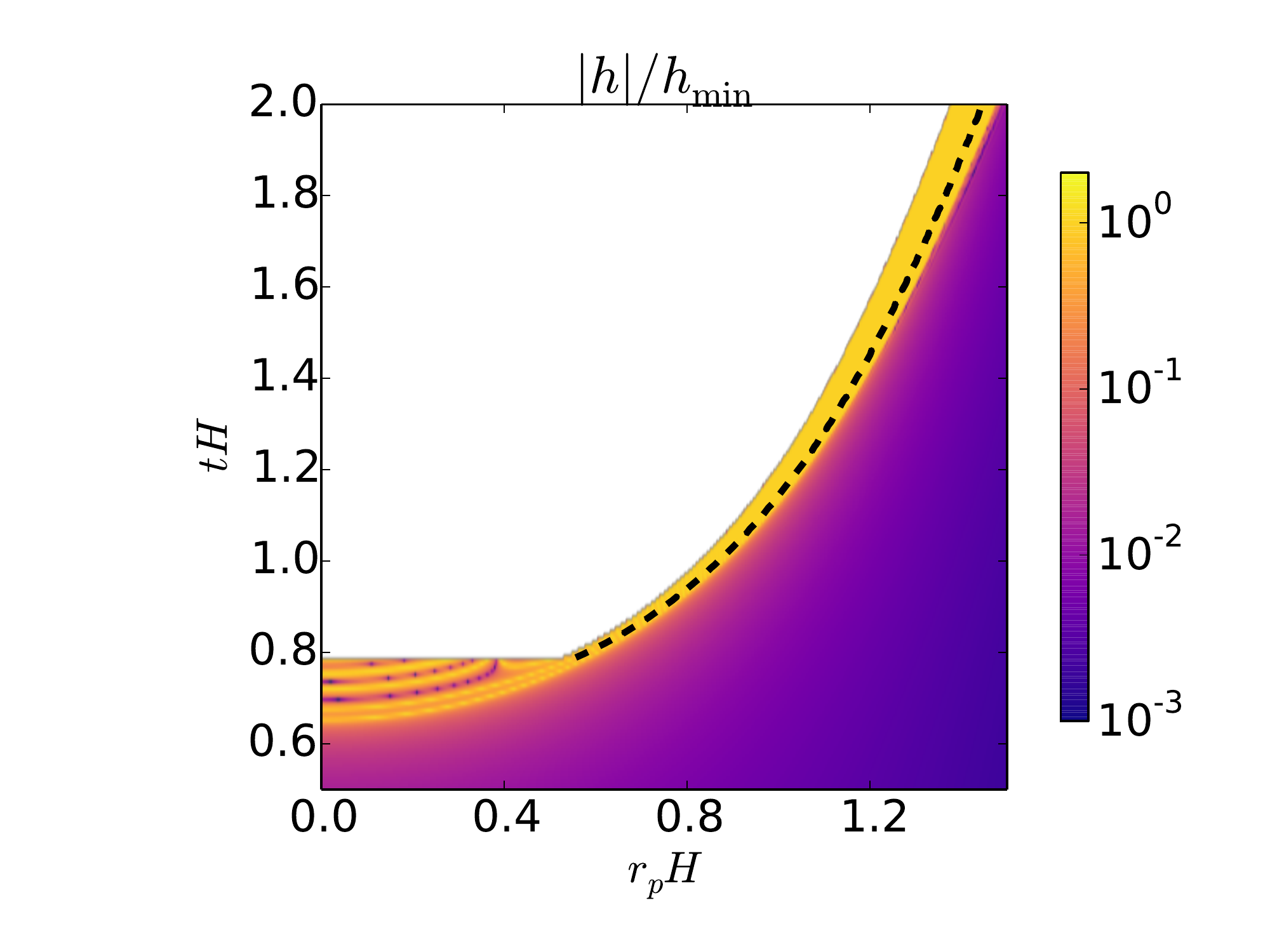}
\hskip-0.3in
\includegraphics[width=0.55\columnwidth,draft=false]{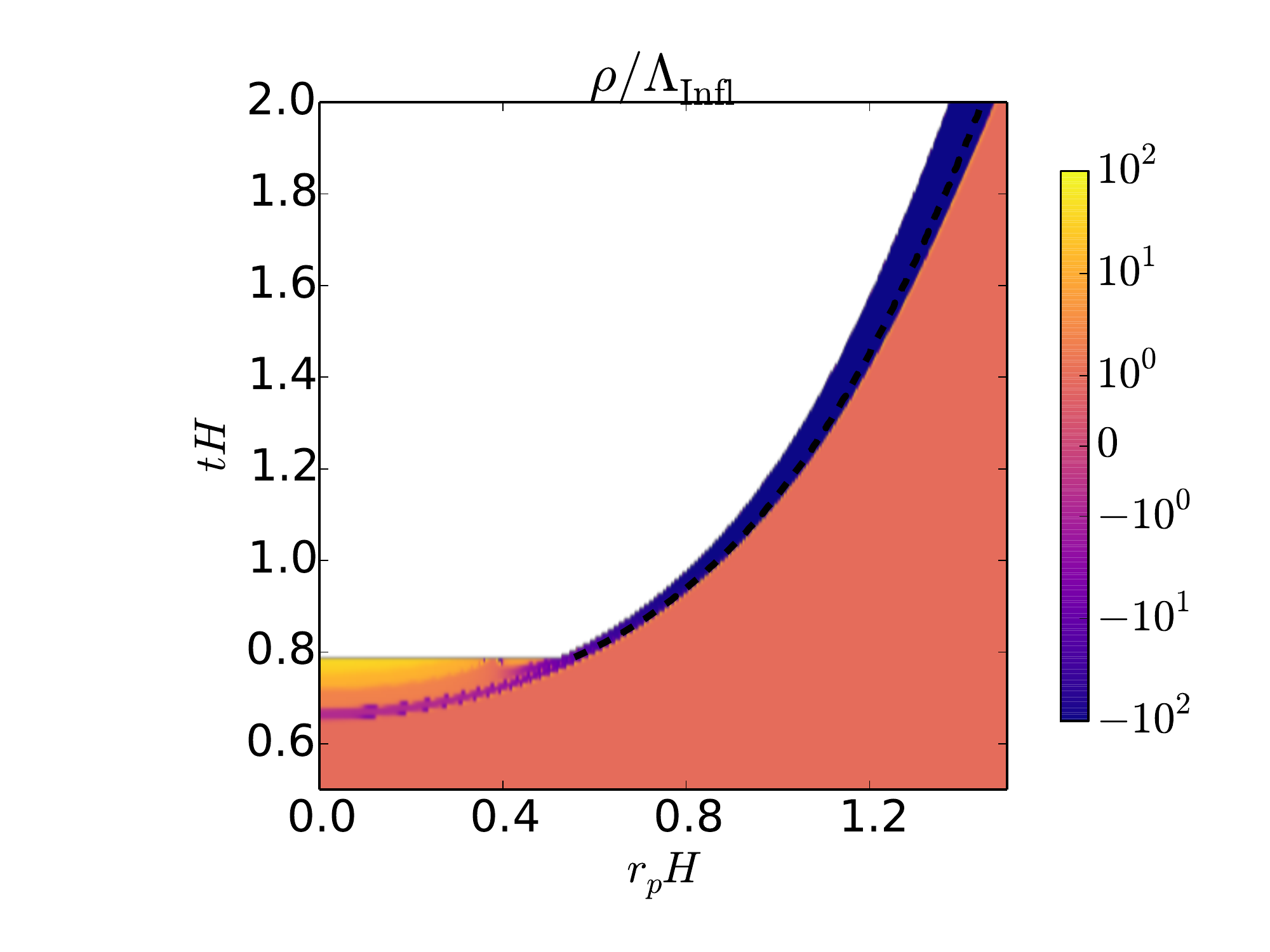}\\
\hskip-0.5in
\includegraphics[width=0.55\columnwidth,draft=false]{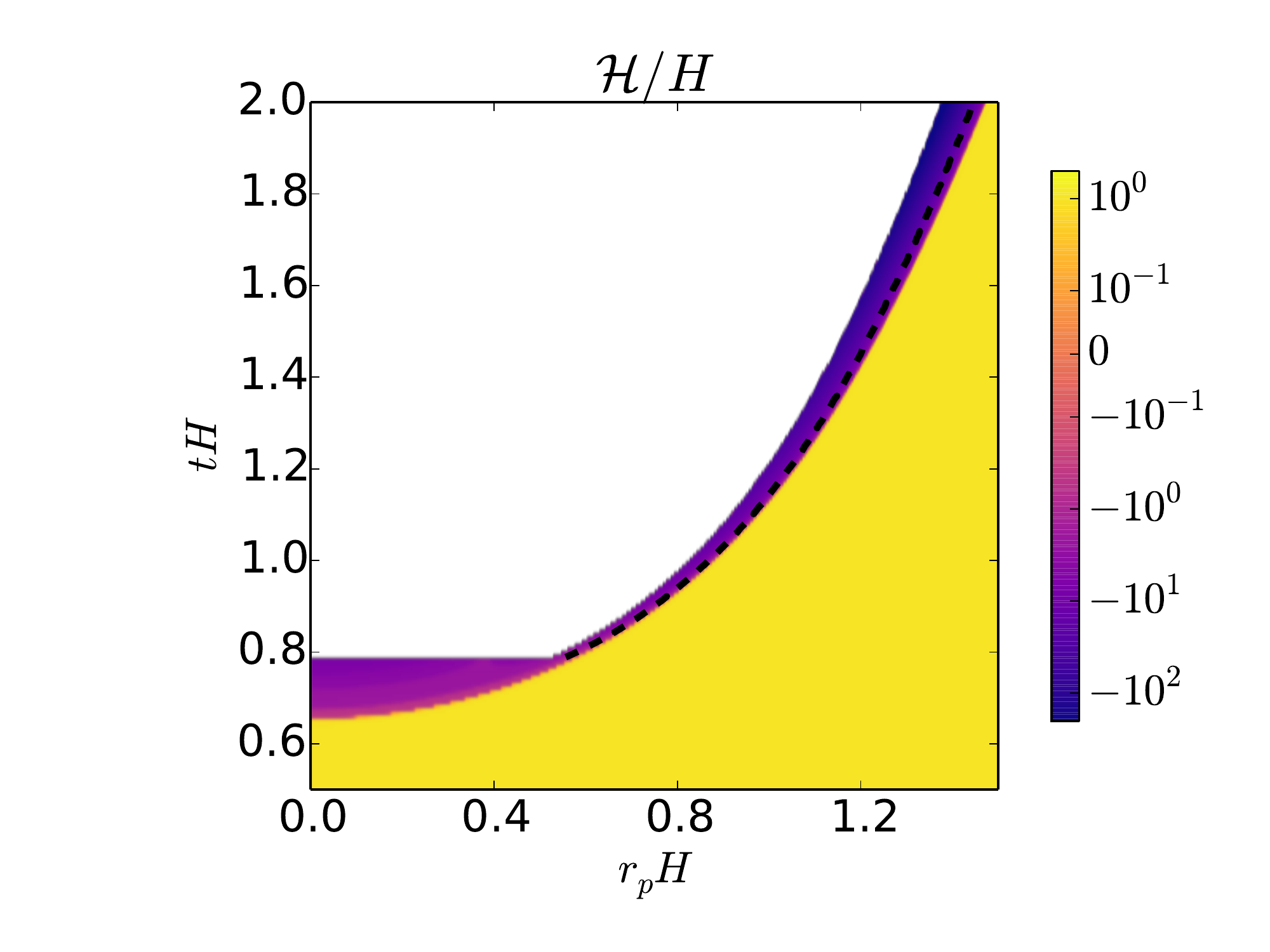}
\hskip-0.3in
\includegraphics[width=0.55\columnwidth,draft=false]{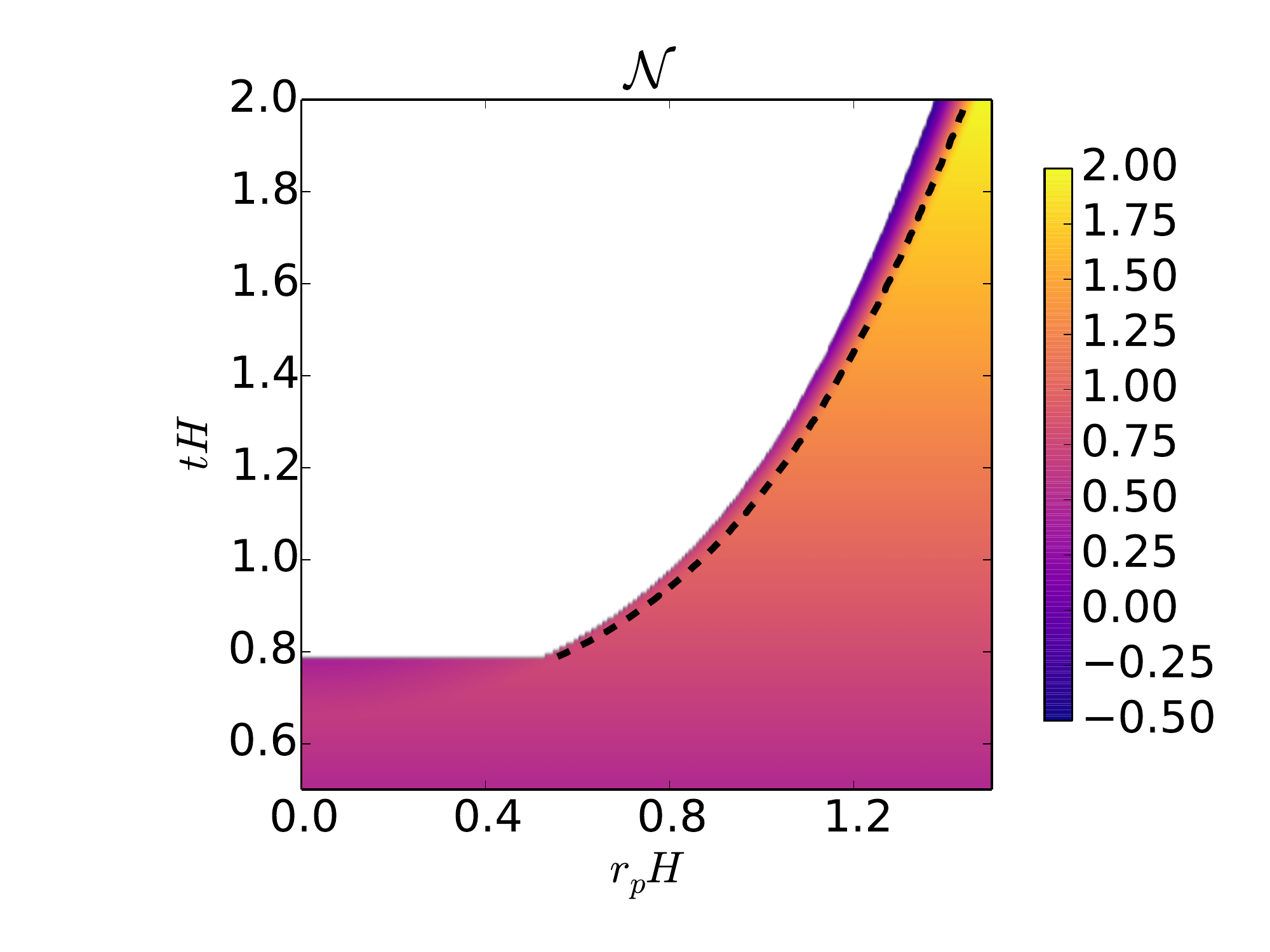}
\end{center}
\caption{
An unstable Higgs field fluctuation falling to the true vacuum. We show, left to
right, top to bottom: the Higgs field value, energy density, local Hubble
expansion rate ($\mathcal{H}=-3K$) and local number of $e$-folds of expansion
as a function of radius (in planar coordinates, $r_p$)  and time. Results correspond to an
initially spatially Gaussian fluctuation with $R=R_H$ and $h_{\rm in}=2 h_{c}$,
and potential with $V_{\rm min}/\Lambda_{\rm Infl}=-100$ and $h_{\rm min}/M_P=0.1$.
The dotted black line indicates the surface of the apparent horizon that forms
during the evolution, while the white space indicates the region behind the apparent horizon that is excised from the domain in order to
continue the simulation. 
\label{fig:bubble} }
\end{figure*}

In \Fref{fig:bubble} we show the evolution of an example case where this occurs (similar results are found for other
parameters). In addition to the value of the Higgs field, we
also plot the energy density $\rho = n_a n_b T^{ab}$, a local measure of the
Hubble expansion rate $\mathcal{H}\equiv \nabla_a n^a/3$, and a local
measure of the number of $e$-folds of expansion $\mathcal{N}$ (found by
integrating $n^a\nabla_a\mathcal{N} = \mathcal{H}$). These quantities are defined
in terms of $n^a$, the unit normal to slices of constant coordinate time. 

Previous studies have ignored the dynamics of the Higgs field in this process;
we find, however, that they are crucial to understanding the evolution
of the system. The Higgs fields falls towards the true vacuum, eventually oscillating
around the global minimum (top-left panel of \Fref{fig:bubble}), with a large
amount of the potential energy liberated by the field going into
kinetic/gradient energy. The energy density at the center of the fluctuation grows
(top-right panel) as this region contracts (bottom-left panel), leading to
the formation of a black hole, as indicated by the presence of an apparent horizon
inside the AdS-like region (denoted by dashed black lines in
\Fref{fig:bubble}). The positive mass of the black hole is compensated by
a shell of negative potential energy surrounding it (see the narrow blue
strip adjacent to the black hole horizon in the top right panel); both increase
in size as evolution progresses, with more and more of the energy obtained
by the Higgs field falling to its true vacuum being locked behind the black
hole horizon.  This black hole hides any crunching singularity (potentially
indicated by the negative values in the bottom-right panel) from outside
observers. Note that the white space in the plots indicates
regions inside the apparent horizon that are excised in order to continue the
simulations. Though (as seen in the top-left panel of \Fref{fig:bubble})
the spatial region over which the Higgs field transitions
from near the true vacuum to a much smaller value is small, as we detail below,
the dynamics of the spreading of the fluctuation is not determined by the 
boundary of the region which has reached the true vacuum, and is thus 
qualitatively different from the thin-wall approximation where all the
evolution is set by this interface. 

\subsection{Dynamics of bubble wall: Causally disconnected evolution}
\label{subsec:bubble}

Next, we study how the fluctuation propagates outwards in spacetime. During
inflation, an unstable Higgs fluctuation will both spread due to dispersion, as
well as expand due to inflation, in spite of the eventual formation of a crunching region in the interior.
The inflationary expansion is more important when the characteristic size of the fluctuation is greater than the Hubble radius, and occurs while the amplitude of the fluctuation increases as it falls to the true vacuum.  

\begin{figure}
\begin{center}
\hskip-0.2in
\includegraphics[width=0.5\columnwidth,draft=false]{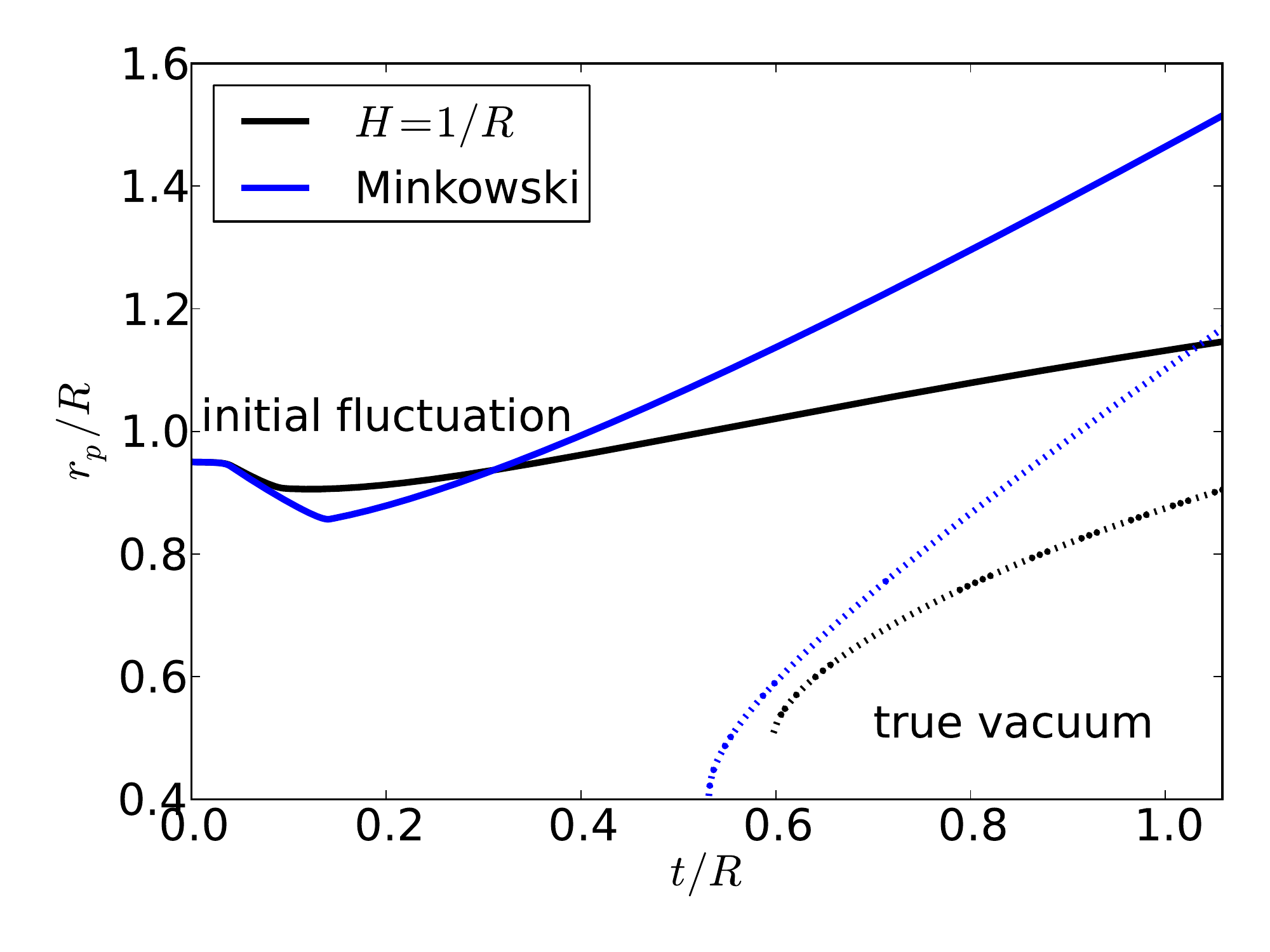}
\includegraphics[width=0.5\columnwidth,draft=false]{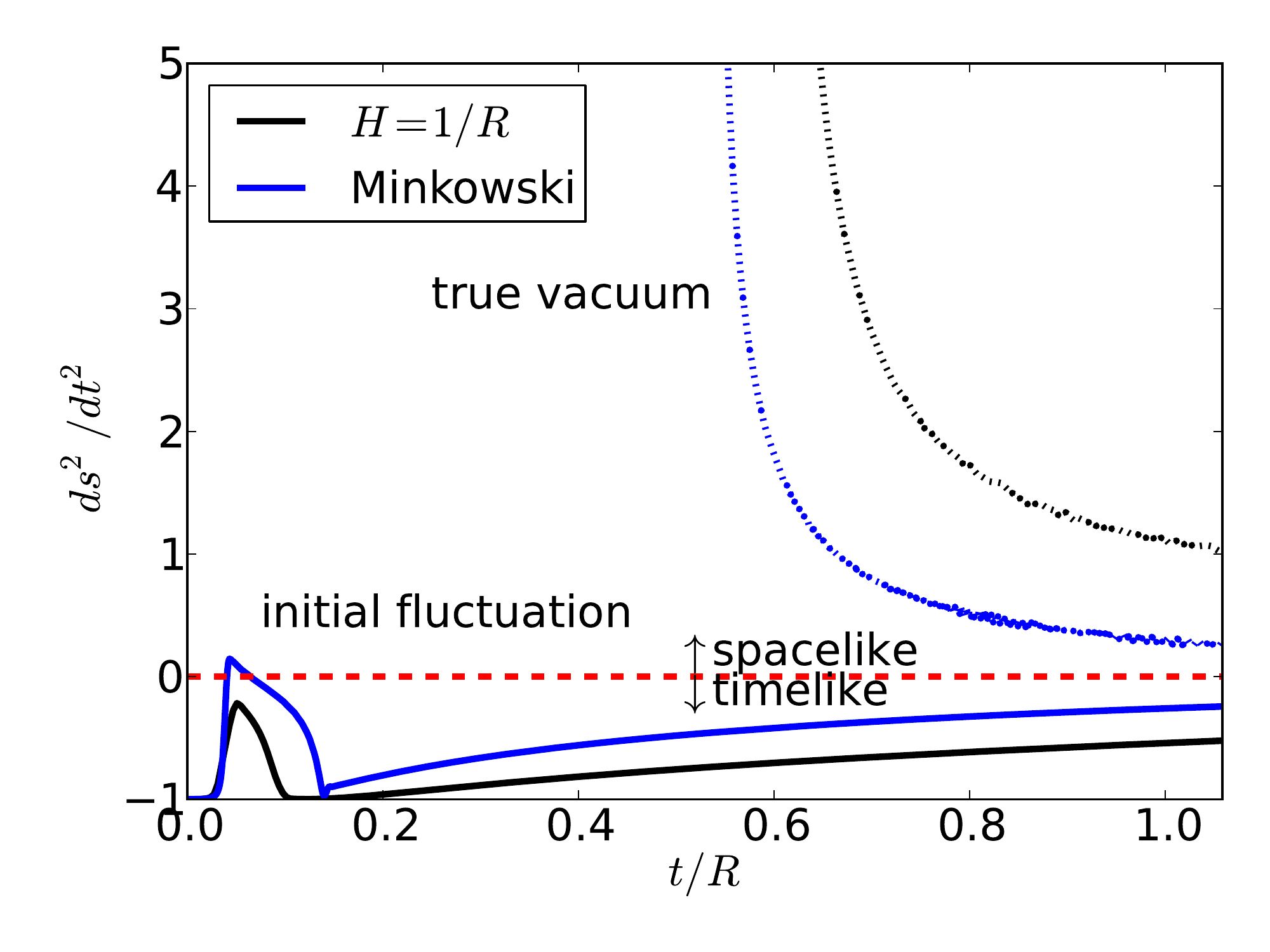}
\end{center}
\caption{ 
Evolution of the size of a large Higgs fluctuation and resulting region of
true vacuum in de Sitter (black lines) and Minkowski (blue lines) space.  The
left panel shows the outermost radius in planar coordinates (hence a factor of
$\exp(tH)$ should be applied to obtain the proper radius), where the Higgs
field equals $h_{\rm in}/2$ (roughly the radius of initial fluctuation;
solid lines) and where the Higgs field equals $h_{\rm min}/2$ (roughly the
radius of the true vacuum patch; dotted lines), as a function of time. The right
panel shows proper length squared per time squared $ds^2/dt^2$ of these curves.
The potential has $V_{\rm min}R^2/3=100$ and $h_{\rm min}=0.1$. The initial
fluctuation has a compactly supported spatial profile given by
\Eref{eqn:step} (the transient behavior at early times being an artifact of this
particular choice).
\label{fig:wall_tophat} }
\end{figure}

For this purpose, two different length scales are of interest: (i) the radius of the unstable Higgs fluctuation (\ie, the region that will rapidly diverge to the true vacuum), defined to be the outermost radius at which the field value is half the amplitude of the initial fluctuation, $h=\frac{1}{2} h_{\rm in}$, 
and (ii) the radius of the bubble of true vacuum (\ie, the region that has effectively reached the global minimum), defined to be the outermost radius where $h=\frac{1}{2}h_{\rm min}$. 
To illustrate the evolution of an unstable Higgs fluctuation and expanding bubble
of true vacuum, in \Fref{fig:wall_tophat} (left panel) we plot how these two
length scales grow as a function of time for a compactly supported Higgs
fluctuation [given by \Eref{eqn:step}].  We plot these for both a Hubble radius-sized fluctuation and the limiting case of Minkowski space.  As expected, in all cases the fluctuation radius increases ahead of the interior expanding bubble of true vacuum.  
While the comoving spread of the fluctuation will be slowed down by Hubble friction in the inflating case relative to Minkowski, its volume will also increase exponentially due to expansion.

The more interesting information is plotted in the second panel of
\Fref{fig:wall_tophat}, which shows the ratio of the change in proper
length squared versus time squared $ds^2/dt^2=g_{ab}(dr^a/dt)(dr^b/dt)$, where
$r^a$ is the spacetime coordinate of the fluctuation or bubble radius. While the edge of the Higgs fluctuation is moving outward at nearly the
speed of light (\ie, $ds^2/dt^2\sim 0$), the growth of the radius of the bubble of true vacuum
is spacelike, $ds^2/dt^2>0$.  This means that one should \emph{not} view
the bubble of true vacuum as causally sweeping outwards, converting dS into AdS.
Rather, the correct interpretation is that after the unstable Higgs fluctuation (at the lower amplitude $\frac{1}{2} h_{\rm in}$ for which it does not backreact on the spacetime)
reaches a given point, that point falls to the true vacuum causally disconnected
from the fact that its neighboring points are also falling to the true vacuum.
In the Minkowski limit, the spacetime curves traced out by the
Higgs fluctuation and the bubble of true vacuum (blue lines) both approach being
null. 
However, in de Sitter space, the exponential expansion eventually
dominates, and the edge of the Higgs fluctuation quickly becomes causally
disconnected from the bubble of true vacuum.  This implies that the growth of the true
vacuum region is insensitive to the behavior of the spacetime in the
interior region and the details of the Higgs potential near $h_{\rm min}$.  
The Minkowski result also illustrates that the true vacuum regions will
continue to grow after the end of inflation when the surrounding energy density
is reduced.

\begin{figure}
\begin{center}
\hskip-0.2in
\includegraphics[width=0.5\columnwidth,draft=false]{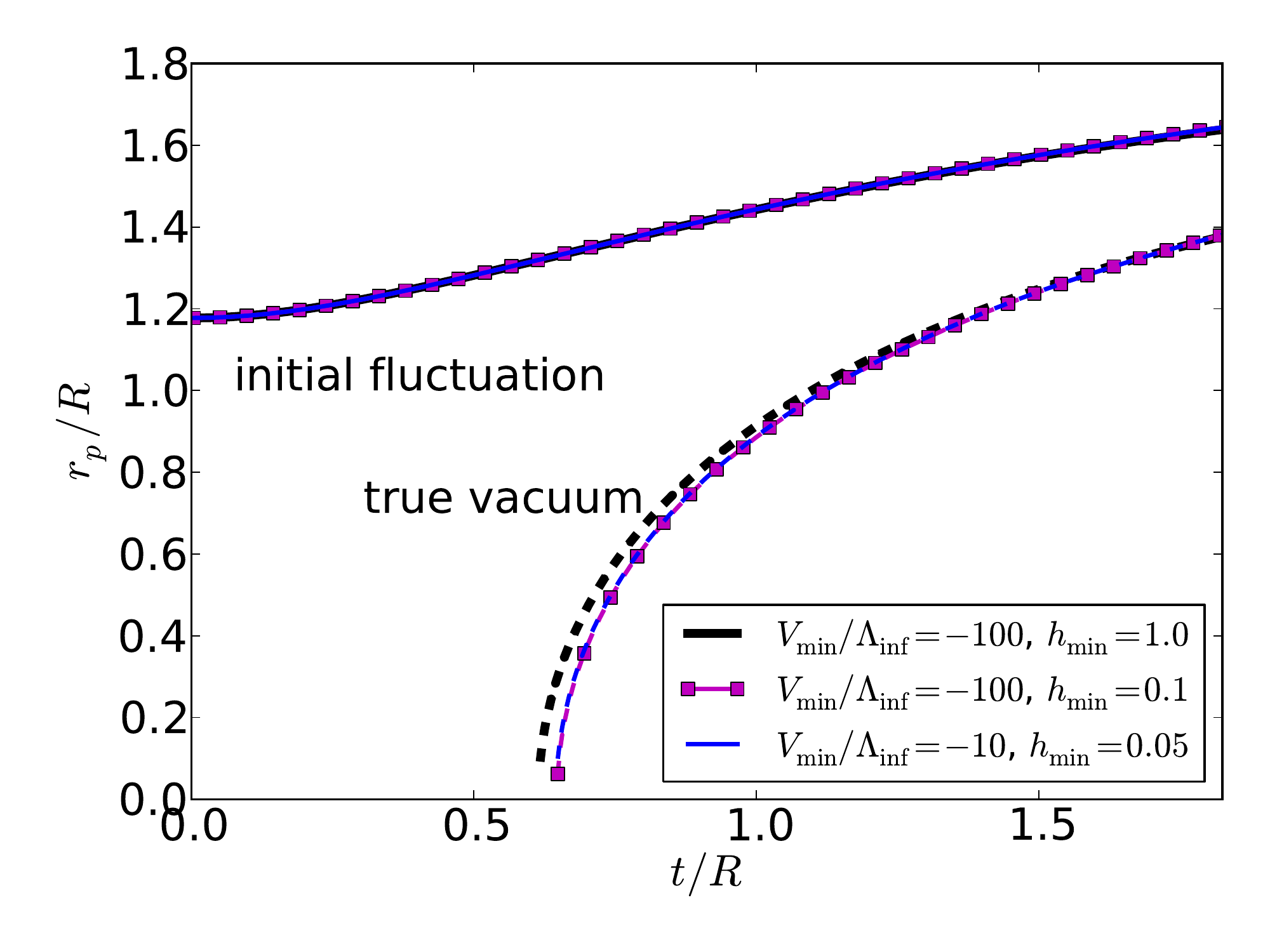}
\includegraphics[width=0.5\columnwidth,draft=false]{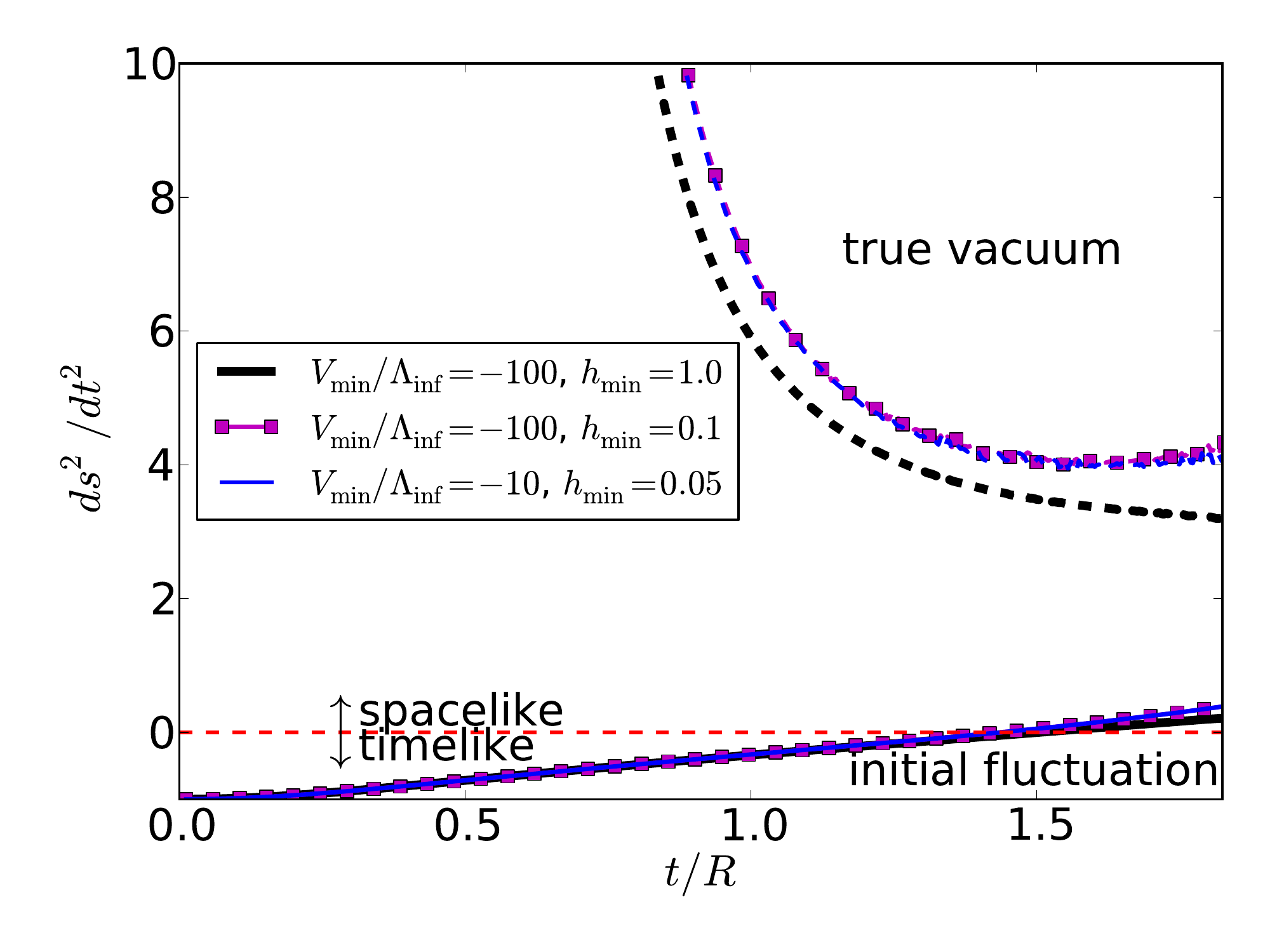}
\end{center}
\caption{
Similar to \Fref{fig:wall_tophat}, but for initially spatially Gaussian
fluctuations in de Sitter with $R=R_H$. The three different cases shown have
$(V_{\rm min}/\Lambda_{\rm Infl}, h_{\rm min}/M_P)=(-100,1.0)$ (black),
(-100, 0.1) (red), and (-10, 0.05) (blue), but show similar behavior.   
\label{fig:wall_gauss} }
\end{figure}

Similar results are also obtained for spatially Gaussian fluctuations, which we show in \Fref{fig:wall_gauss}. It should be noted that the location of the boundary of the fluctuation is less well defined in this case (and the boundary region is also being expanded out of casual contact as seen in the left panel at late times).   
This plot demonstrates that the exact
parameters of the Higgs potential near its minimum make little difference to
the growth. Finally, we note that while a region of true vacuum can become
exponentially large during the de Sitter phase, it of course cannot extend past
the cosmological horizon of the initial Higgs fluctuation that gave rise to it. 
So, the creation of a single unstable fluctuation cannot globally terminate
inflation. However, this may occur if a sufficient proportion of the space
transitions to a crunching phase \cite{Sekino:2010vc}, perhaps implying
constraints on any phase of inflation occurring before that giving rise to our observable Universe.

\subsection{Beyond spherical symmetry: Black holes and violation of hoop conjecture }
\label{subsec:nonspherical}

So far, we have focused on spherically symmetric configurations. However, the assumption
of spherical symmetry strongly limits the spacetime dynamics (for example
precluding the existence of gravitational waves) and hence the range of
solutions uncovered by our simulations. Furthermore, since a large Higgs
fluctuation that has become classical arises from the stochastic addition of
many modes, there is no reason to expect it to be spherical, so that it is
important to study this broader class of fluctuations. 
For these reasons, we now extend our studies to nonspherically symmetric
(though still axisymmetric) cases.

\begin{figure}
\begin{center}
\hskip-0.5in
\includegraphics[width=0.55\columnwidth,draft=false]{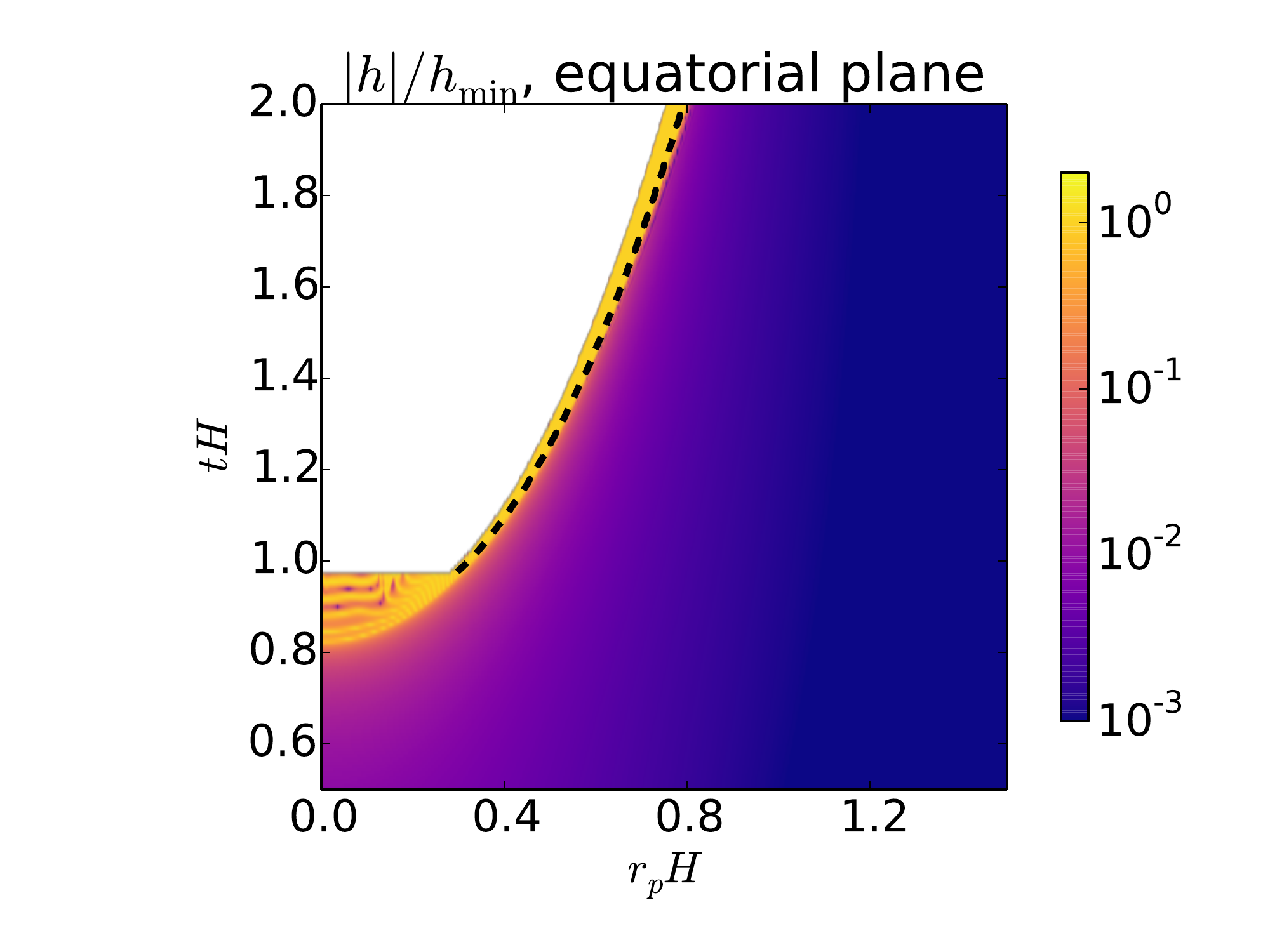}
\hskip-0.3in
\includegraphics[width=0.55\columnwidth,draft=false]{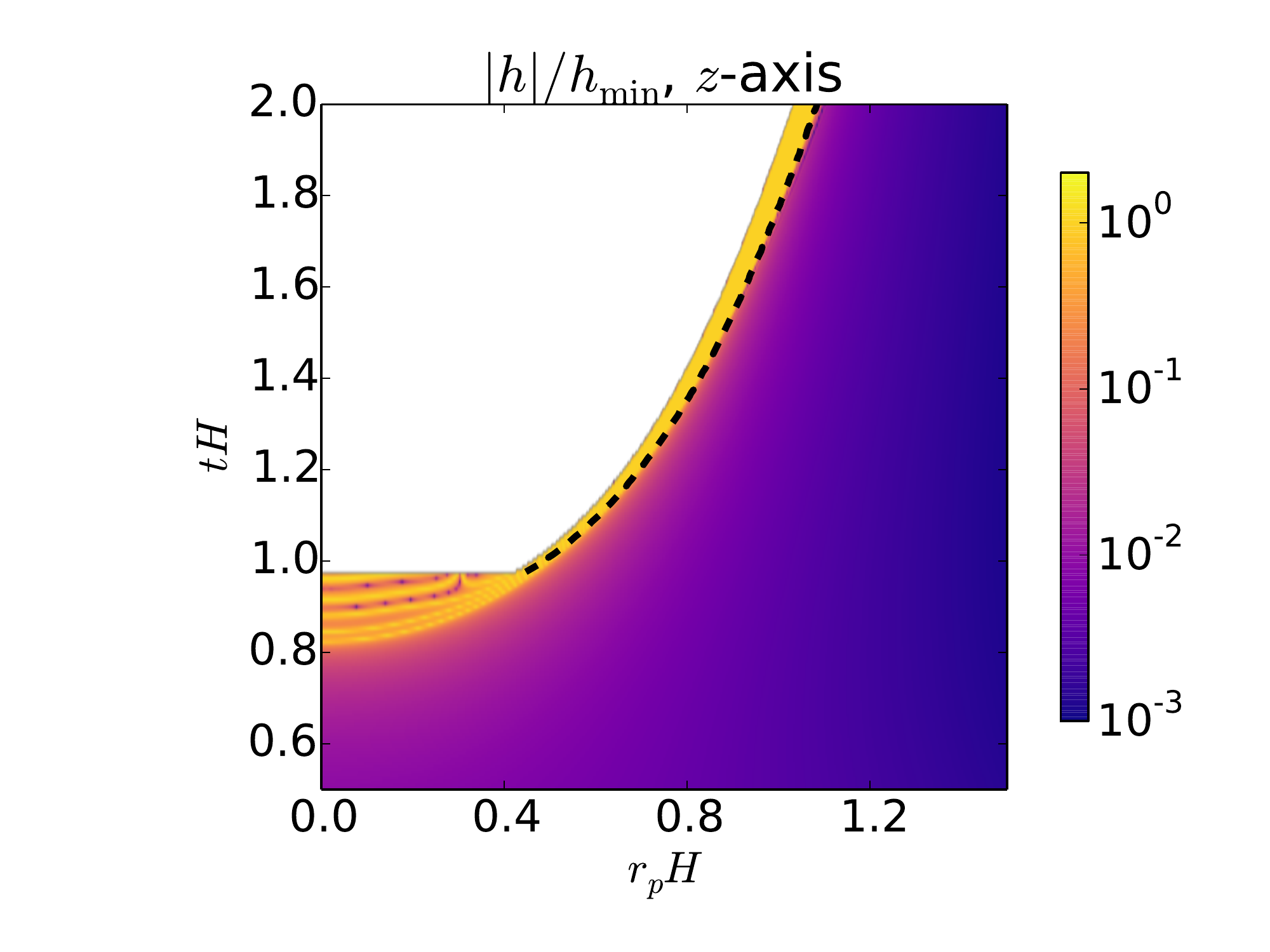}
\end{center}
\caption{
The Higgs field fluctuations as a function of radius (in planar coordinates, $r_p$) and time for the evolution of an initial
fluctuation that is an elliptical Gaussian with $R_z=2R_{xy}=R_H$.  The left
panel shows the field in the equatorial plane, while the right panel shows the
field on the symmetry axis.  Except for the absence of spherical symmetry, the
parameters in this case are the same as the ones shown in \Fref{fig:bubble},
and the evolution proceeds in a similar manner.
\label{fig:ellip_bubble} 
}
\end{figure}

We find that such configurations evolve similarly in many ways to the
spherically symmetric cases considered above. In \Fref{fig:ellip_bubble}, we
show results from a case identical to the one shown in \Fref{fig:bubble},
except with $R_{z}=2R_{xy}=R_{H}$ instead of $R_{xy}=R_{z}=R_{H}$. 
In both cases, the field swiftly falls to the true vacuum, creates a
crunching region, and forms an apparent horizon.  Thus our observations from
the previous subsections also apply to nonspherical configurations.

More interestingly, we find that there are significant differences between the
large Higgs fluctuation cases we study here---which produce regions of
negative energy density---and spacetimes that satisfy standard energy
conditions. In particular, for four-dimensional spacetimes with positive
energy, black hole apparent horizons are found to always have spherical
topology~\cite{hawking1973large,1994PhRvD..49.6467H}. Furthermore, it has been
found that, geometrically, black holes cannot be arbitrarily elongated.
The latter condition is encapsulated in the hoop conjecture~\cite{thorne_hoop},
which states that a region containing a mass $M$ will form a black hole with
attendant horizon if and only if a ``hoop" of circumference $4\pi M$ can be
passed over the region in every direction. For example, the collapse of an
infinite cylinder will not form a horizon, but instead create a naked
singularity.  Crucially, these restrictions do not apply to AdS spacetimes,
which can develop cylindrical black holes~\cite{Lemos:1994xp,Lemos:1997bd}. Analogously, we
find that they also do not apply in our study of Higgs fluctuations because
the regions in which the Higgs field diverges to the true vacuum evolve into regions
with negative potential energy, allowing for the formation of arbitrarily elongated black holes. 

\begin{figure}
\begin{center}
\includegraphics[width=0.45\columnwidth,draft=false]{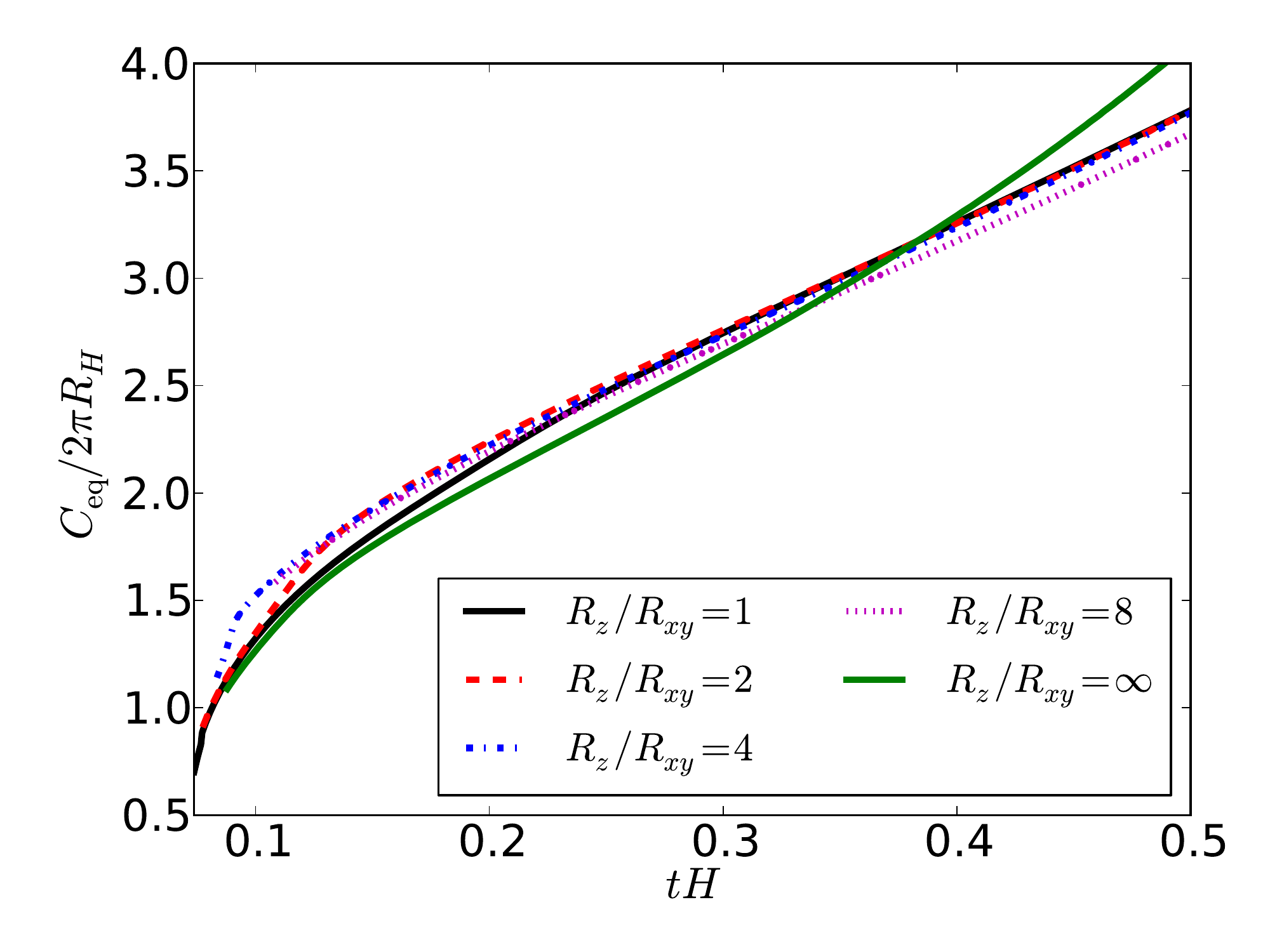}
\includegraphics[width=0.45\columnwidth,draft=false]{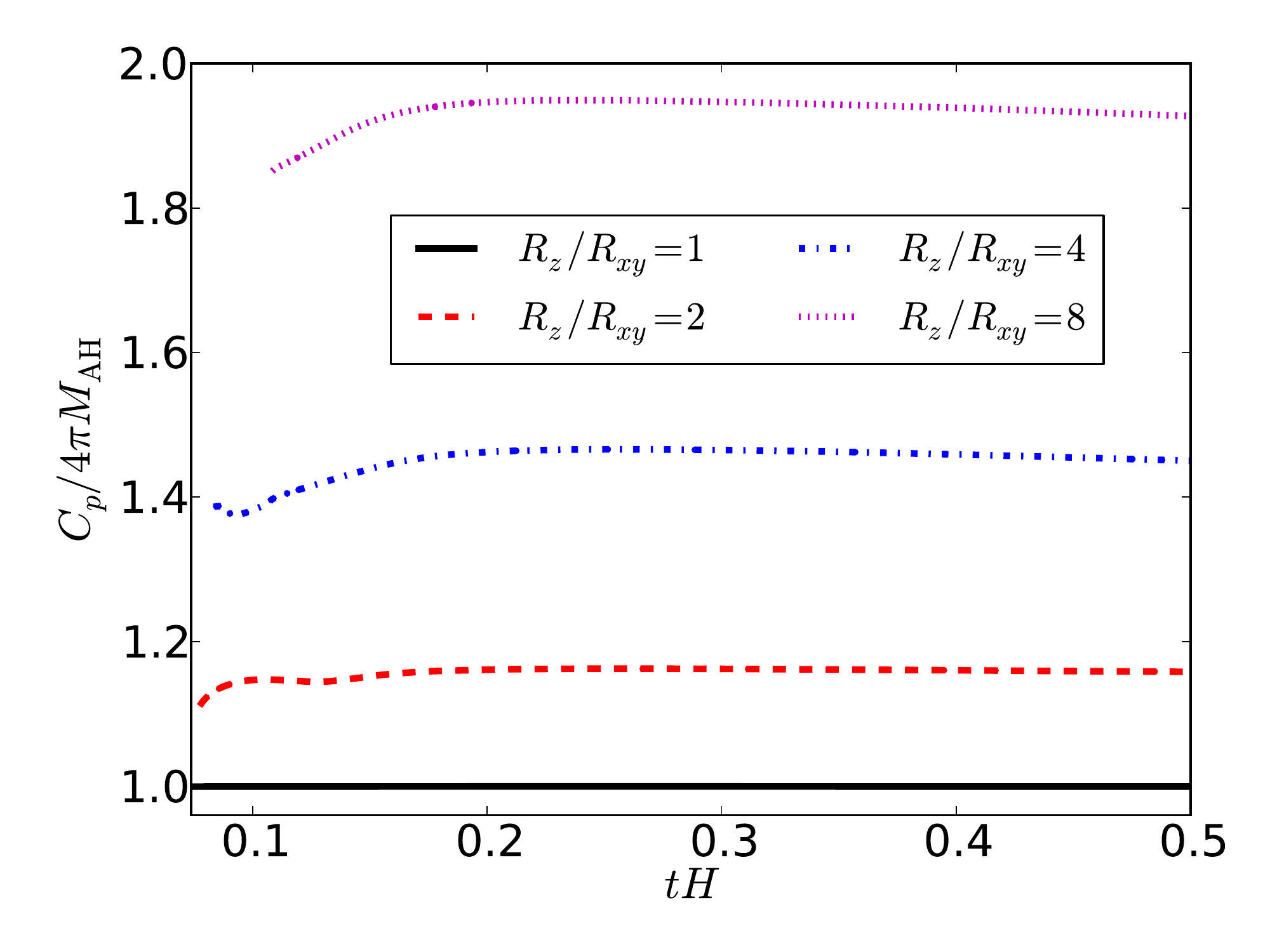}
\end{center}
\caption{
  The proper equatorial (left panel) and poloidal (right panel) circumferences
  of the apparent horizons that form from large Higgs field fluctuations
  (beginning from when they are first found in the domain) with various aspect
  ratios $R_z/R_{xy}$. We also show the equatorial circumference of a case with
  $z$-translational symmetry ($R_z/R_{xy}=\infty$) where the horizon has
  cylindrical topology.  The poloidal circumference is normalized by the horizon
  mass $M_{AH}$ to show how far above the hoop conjecture criterion $C\lsim 4
  \pi M_{\rm AH}$ it is in each case.
\label{fig:long_bhs} 
}
\end{figure}

To demonstrate this, we consider a series of increasingly elongated Gaussian
field configurations.  We fix the radius in the equatorial plane, $R_{xy}=R_H$,
and consider cases with larger and larger extent along the symmetry axis,
$R_z/R_{xy}=1$, 2, 4, and 8.  In all cases we find that an apparent horizon
does form soon after the Higgs fluctuation reaches the true vacuum.  As shown in
\Fref{fig:long_bhs}, the proper equatorial circumference $C_{eq}$ of the
horizon evolves in a similar manner for all cases, indicating that the narrow
``waist" of the Higgs fluctuation and resulting black hole is not sensitive to
the longer direction.  The poloidal circumference $C_p$ does, however, increase with 
the increasing aspect ratio, and it increases at a greater rate than the mass of the
horizon $M_{\rm AH}$ (measured from its proper area), giving larger and larger
violations of the hoop conjecture criterion. 

Although configurations with larger aspect ratios become increasingly difficult
to track numerically, we can also consider the infinite $R_z/R_{xy}$ case by
evolving a spacetime with an enforced translational symmetry in the
$z$-direction.\footnote{For this one case, we use periodic boundary conditions
in the $x$ and $y$ directions, though with the boundary at large enough
distances so as to be insignificant to the results shown here.}
We find that a horizon also forms in this case, now with
\emph{cylindrical} topology, \ie, a black string.  From \Fref{fig:long_bhs}
we can see that the evolution of its circumference proceeds in a similar manner
to the other cases.  These results suggest that, even proceeding from a slice of
$3+1$ dimensional (very nearly) de Sitter spacetime, the presence of a Higgs
field potential with a negative minimum enables the formation of arbitrarily
elongated, stringlike horizons. 
Thus the hoop conjecture is violated in favor of ensuring that cosmic censorship---the requirement that singular regions be hidden from outside observers---is obeyed.

\section{Numerical Solutions of the Fokker-Planck Equation and Implications for Cosmological History}
\label{fp_limits}

In the previous section, we established with solutions in full General Relativity
that, for reasonable parameters, regions of space exhibiting sufficiently large
Higgs fluctuations $\abs{h} \gsim \hsrb$ do rapidly fall to the true vacuum and create
persistent crunching regions, both in an inflating spacetime, and in the
limiting case of an approximately Minkowski spacetime. The formation of any such
fluctuation during inflation would therefore be fatal for our Universe, as it
would expand after inflation has ended, destroying the surrounding
(approximately) Minkowski space in the EW vacuum. Having established this
result, we now return to the stochastic approach, introduced in
\Sref{evolution_stages}, to determine the implications for the scale of
inflation.  In contrast to previous work that also solved the FP equation, here we numerically resolve the exponentially suppressed tails of the distribution, which, as we will show, impacts the constraint on $H/\Lambda_{\rm max}$.  Below, in \Sref{subsec:approximate}, we compare the exact solution to previous approximations.

We are interested in solving the FP equation to determine when at least one true
vacuum patch has formed in our past light cone, \ie,
\be
P(\abs{h} \gsim \hsrb,N) e^{3N} \geq 1
\label{Eq:OnePatch}
\ee
after $N$ $e$-folds of inflation. We rewrite the FP equation (\Eref{Eq:FP}) in terms of the variable $X \equiv \log P$,
\be
\frac{\partial X}{\partial t} = \frac{V''(h)}{3H} + \frac{V'(h)}{3H} \frac{\partial X}{\partial h} + \frac{H^3}{8\pi^2} \left[\frac{\partial^2 X}{\partial  h^2} + \left(\frac{\partial X}{\partial h}\right)^2\right],
\label{eqn:FPX}
\ee
and numerically solve for $X$.  This is essential for resolving the exponentially small tails of the distribution that determine probabilities, of order $e^{-3N}$, for obtaining $\abs{h} \gsim \hsrb$.
Given $X$, one can calculate, for instance, the maximum number of $e$-folds $N_{\rm max}$ that inflation can proceed without the formation of a true vacuum patch, \ie, without \Eref{Eq:OnePatch} being satisfied.
Note that, as stressed in \cite{Espinosa:2015qea}, integrated ``transition probabilities'' to find the Higgs field beyond a certain value, such as $P(\abs{h} \gsim \hsrb,N)$ in \Eref{Eq:OnePatch}, are gauge invariant. However, for simplicity and to allow comparison to prior results, we will frequently quote results in terms of $\Lmax$ (or rather $H/\Lmax$) computed in Landau gauge.

As we are interested in the formation of a true vacuum patch only in our past light cone, we need only consider the ensemble of patches arising from the progenitor patch that inflated $N \simeq 50$--$60$ $e$-folds before the end of inflation to give rise to our observable Universe \cite{Liddle:2003as}. We (optimistically) assume this patch began the relevant period of inflation in the electroweak vacuum, \ie, with $P(h,0) \simeq \delta(h)$.
More details on solving the FP equation, including the exact initial conditions used, are given in the Appendix.

We calculate the appropriate Higgs quartic for \Eref{eq:VhforFP} using two-loop
renormalization-group improved couplings and including one-loop contributions to
the effective potential \cite{Cheung:2012nb}, specifically
\be
V_{\rm eff} = \frac{\lambda(\mu) + \lambda^{(1)}_{\rm eff}(\mu \simeq \sqrt{h^2 + H^2})}{4} h^4
\ee
where the one-loop contribution to the quartic is
\begin{align}
\lambda^{(1)}_{\rm eff}(\mu \simeq \sqrt{h^2 + H^2})) & \simeq \frac{1}{16 \pi^2} \left\{\frac{3 g_2^2}{8} \left(\log \frac{g_2^2}{4} - \frac{5}{6}\right) + \frac{3 (g_2^2 + g_Y^2)}{16} \left(\log \frac{g_2^2 + g_Y^2}{4} - \frac{5}{6}\right)\right. \nonumber \\ & \qquad \qquad \left. - 3 y_t^4 \left(\log \frac{y_t^2}{2} - \frac{3}{2}\right)\right\}.
\end{align}
We match observed
quantities to $\overline{\rm{MS}}$ parameters using expressions from
\cite{Buttazzo:2013uya}.  In the parameter space of interest, we find that the scale at which the quartic is to be evaluated, $\mu \simeq \sqrt{H^2 + h^2}$, lies within approximately an order of magnitude of
$\Lmax$. As such, a suitable approximation for the running coupling is
\be
V(h) \simeq - b_0 \log\left(\frac{H^2 + h^2}{\sqrt{e} \Lmax^2}\right) \frac{h^4}{4}.
\label{eq:higgspot}
\ee
Taking the central values for the Higgs mass $m_h = 125.09 \, \pm \, 0.24 \GeV$ \cite{Aad:2015zhl}
and the top quark mass $m_t = 172.44 \, \pm 0.70 \, \GeV$ \cite{Khachatryan:2015hba}, we find $b_0
\simeq \frac{0.12}{(4 \pi)^2}$ and $\Lmax \simeq 3.0 \times 10^{11}
\GeV$.\footnote{Here, we use the recently updated measurement of $m_t$ from CMS
as it is subject to the smallest uncertainties, but note that ATLAS has also recently
published a comparable measurement $m_t = 172.84 \pm 0.86 \GeV$
\cite{Aaboud:2016igd}. In addition to the experimental uncertainties, we have
added in quadrature $0.5 \GeV$ of theoretical uncertainty to account for
conversion between a Monte Carlo and on-shell top mass definition
\cite{Hoang:2008xm,Moch:2014tta}.}
The corresponding values of $\hcld, \hsrb$ depend somewhat on $H$, but tend to be $\hcld \simeq 1.2 \Lmax$ and $\hsrb \simeq {\rm few} \times \Lmax$ in the parameter space of interest.

\begin{figure}
\includegraphics[width=0.48\columnwidth]{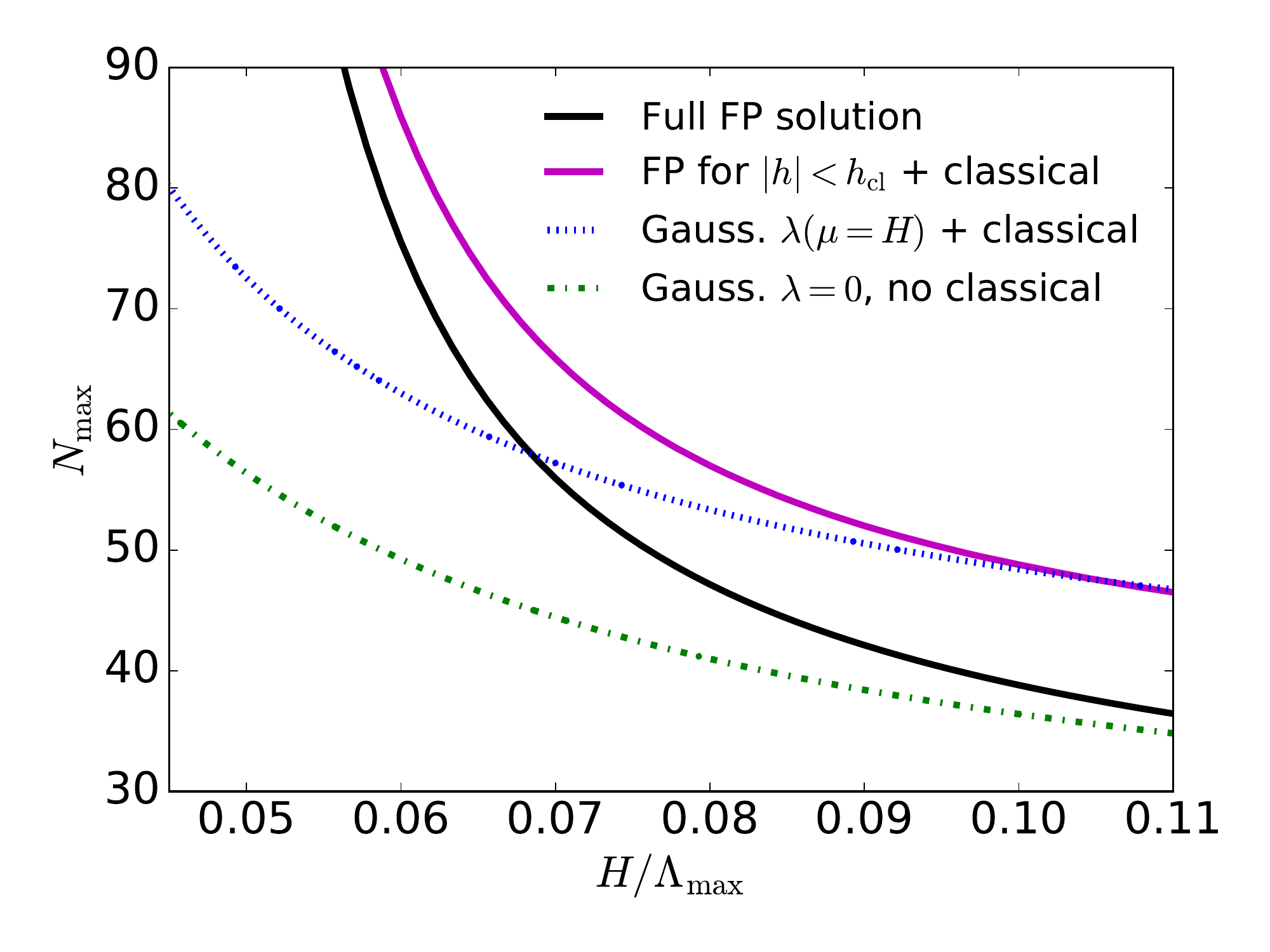}
\includegraphics[width=0.48\columnwidth]{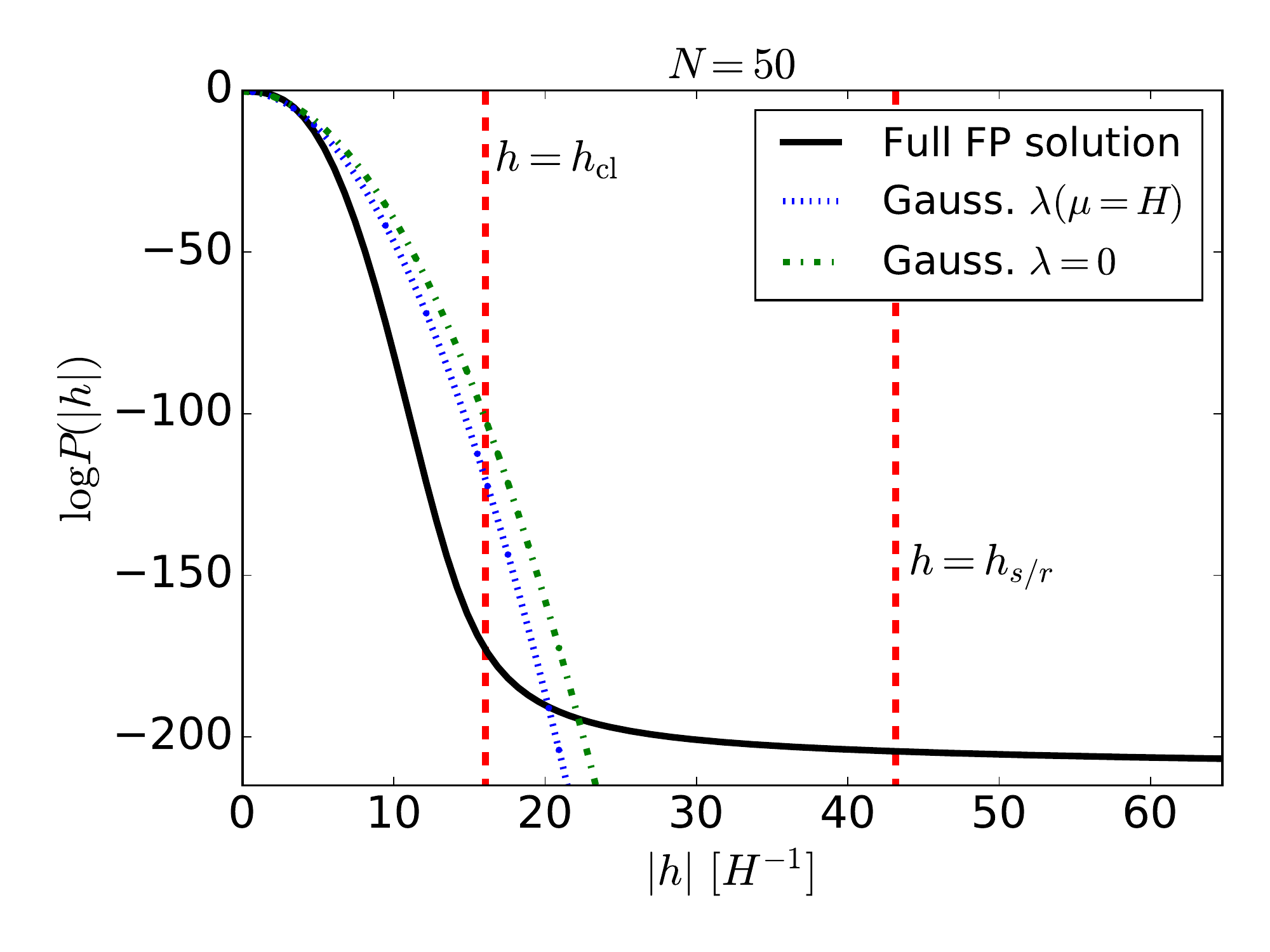}
\caption{Left panel: The maximum number of
$e$-folds that inflation can proceed without the formation of a true vacuum
patch $N_{\rm{max}}$ as a function of $H/\Lambda_{\rm{max}}$.  Right panel:
Probability distribution of the Higgs field after $N = 50$ $e$-folds for
$H/\Lmax = 0.067$. In both cases, the full FP treatment is compared to other
approaches. Note the long, non-Gaussian tail that develops at high field values 
in the right panel due to (as discussed below) the strong effect of the negative
quartic in this regime. 
}
\label{fig:Nelimit}
\end{figure}

The dependence of $N_{\rm max}$ on $H/\Lmax$ is shown in the left panel of
\Fref{fig:Nelimit} (solid, black). In particular, if we require that inflation
lasts at least 60 $e$-folds, we find
\be
\label{eq:conservativelimit}
\boxed{\frac{H}{\Lmax} \lsim 0.067 \; \Rightarrow \; \mbox{no true vacuum patches (\ie, with $\abs{h} > \hsrb$) form during inflation}}
\ee
for the central values of $(m_h, m_t)$ quoted above.

We note that this limit is maximally conservative---for $H/\Lmax$ satisfying
\Eref{eq:conservativelimit}, patches in which $\abs{h} > \Lmax$ may still be
formed. These can in principle be stabilized by, \eg, efficient reheating \cite{Espinosa:2015qea}, but this implies a condition on postinflationary
cosmology. If reheating is not sufficiently efficient to drive these patches
back to the electroweak vacuum, they will ultimately classically evolve to the
true vacuum, which would still prove disastrous for our Universe. Thus, we can
also consider the more stringent requirement that no patches in our past
light cone fluctuate beyond the maximum of the potential during inflation.
In this case, we find
\be
\label{eq:nostablelimit}
\boxed{\frac{H}{\Lmax} \lsim 0.064 \; \Rightarrow \; \mbox{no patches with $\abs{h} > \Lmax$ form during inflation}.}
\ee
These are our main results, and represent the most accurate constraints on $H$ in the presence of a SM vacuum instability.

\begin{figure}
\includegraphics[width=0.7\columnwidth]{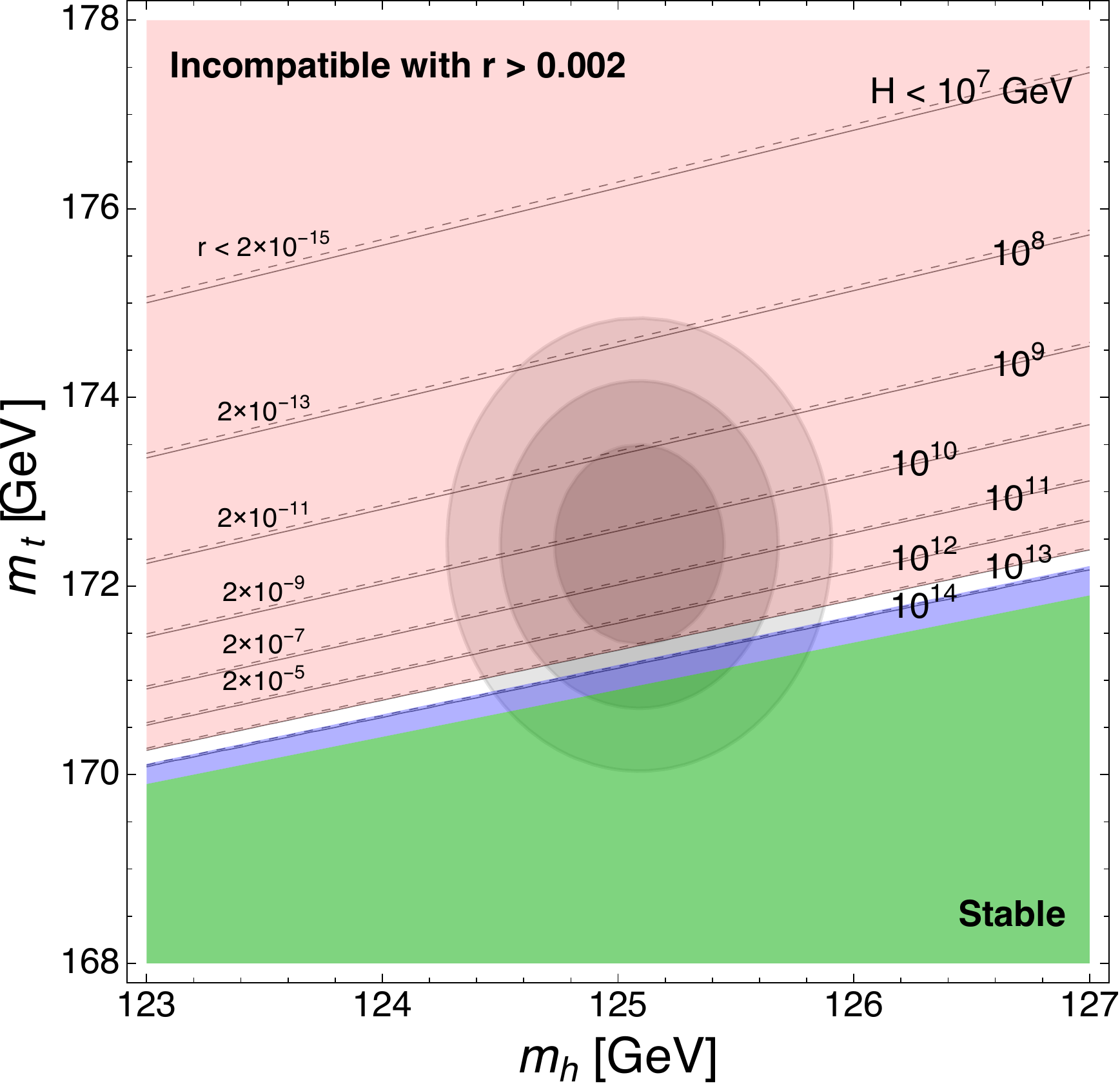}
\caption{Limits on $H$ [GeV] (black contours) in the $(m_h,m_t)$ plane requiring
$N_{\rm max} \geq 60$ (solid) or $50$ (dashed). Central values are taken to be
$m_h = 125.09 \pm 0.24 \GeV$ \cite{Aad:2015zhl} and $m_t = 172.44 \pm 0.70 \GeV$
\cite{Khachatryan:2015hba}, with contours corresponding to 1-, 2-, and
3-$\sigma$ regions as for two parameters. The shaded regions represent: the Higgs
potential is stable up to $M_P$ (green); the Higgs potential is unstable, but current
limits $r < 0.07$ \cite{Array:2015xqh} permit required amount of inflation
(blue); and instability would preclude the combination of $N_{\rm max} > 60$ and
$r > 0.002$, to be probed by near-future experiments \cite{Creminelli:2015oda}
(red).}
\label{fig:HIlimit}
\end{figure}

\begin{figure}
\begin{center}
\includegraphics[width=0.55\columnwidth,draft=false]{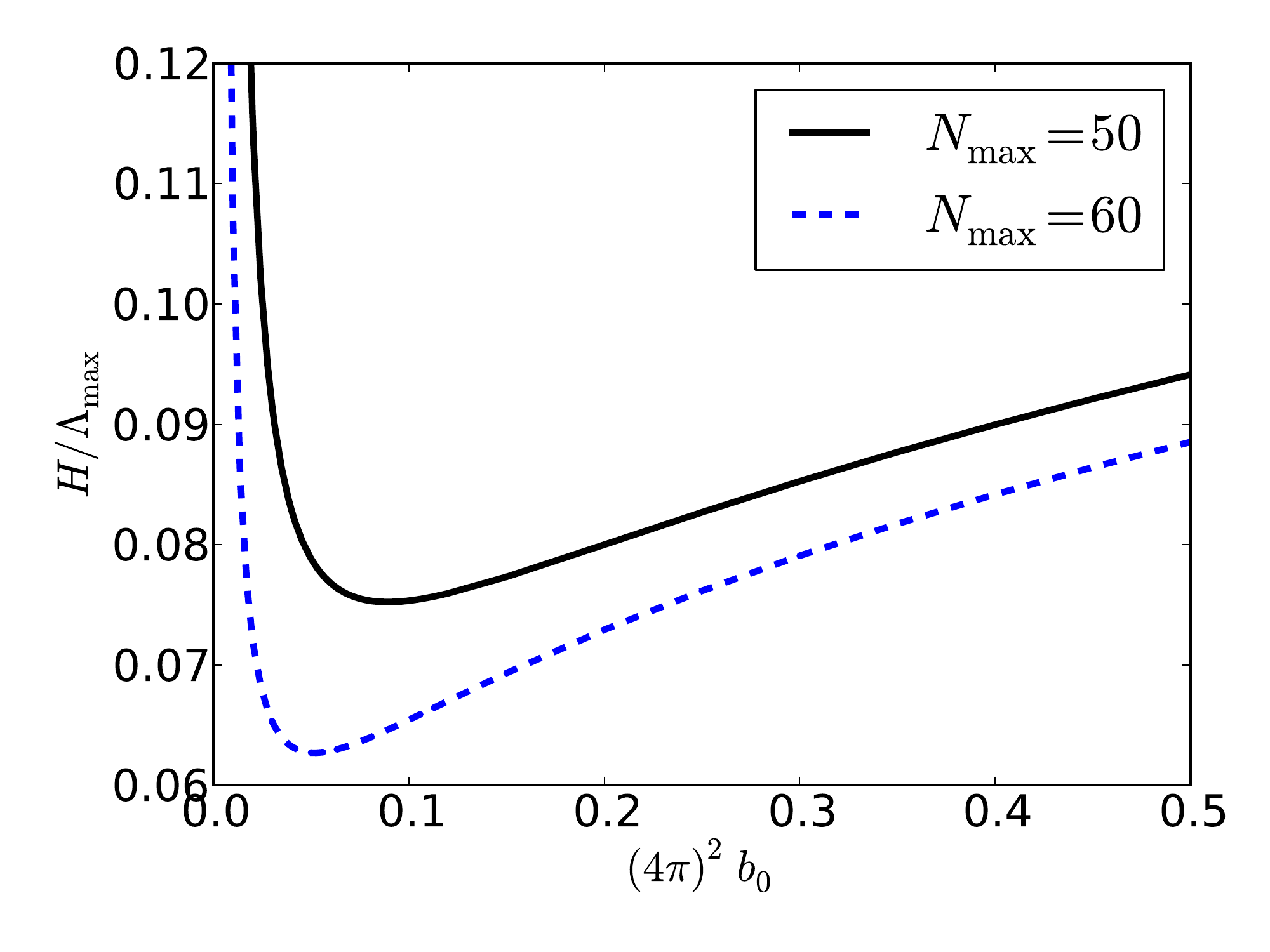}
\end{center}
\caption{
The bounds on the ratio $H/\Lambda_{\rm max}$ as a function of $b_0$, the
$\beta$ function near the maximum of the potential (see \Eref{eq:higgspot}),
requiring $N_{\rm max}=50$ (black, solid) or $N_{\rm max}=60$ (blue, dashed).
\label{fig:HLmaxb0}
}
\end{figure}

We present these results in the $(m_h, m_t)$ plane in \Fref{fig:HIlimit}, taking
the maximally conservative upper limit on the inflationary Hubble scale subject
to the requirement $N_{\rm max} \geq 60$ (solid) or $N_{\rm max} \geq 50$
(dashed).  The limit on $H/\Lmax$ varies nontrivially with $b_0$, as can
be seen in \Fref{fig:HLmaxb0}. 
For fixed $H/\Lmax$, larger $b_0$ (corresponding to larger $m_t$ for a given $m_h$) results in a more positive quartic at the scale $\mu \simeq H < \Lmax$ relevant for small fluctuations, which produces a greater stabilizing effect. But, it also leads to more rapid growth of larger (superbarrier) fluctuations for which the quartic is more negative.
As such, the variation
in the limit depends on which effect dominates. Interestingly, the limit is
approximately strongest for the value of $b_0$ favored by the central
$(m_h,m_t)$ values. However, this limit depends only weakly on $b_0$
throughout the SM parameter space, ranging between $0.06 \lsim
\frac{H}{\Lmax} \lsim 0.11$ for $\frac{0.01}{(4 \pi)^2} \lsim b_0 \lsim
\frac{0.40}{(4 \pi)^2}$ and $N_{\rm max}=60$.  Hence, the bounds on $H$
are mainly driven by how $\Lmax$ varies with $(m_h, m_t)$ and, for a given
$m_h$, the rapid decrease in $\Lmax$ with increasing $m_t$ results in
significantly more stringent limits on $H$.

The region in which the Higgs potential is stable up to the Planck scale is
shown in green, while the region in blue corresponds to where the potential is
unstable, but current limits on $r < 0.07$ \cite{Array:2015xqh} permit the Universe to exit inflation without producing a patch of true vacuum. We also highlight in red the parameter space where the vacuum instability would preclude $N_{\rm max} > 60$ for $r \gsim 0.002$. In other words, in the event of a near-future detection of primordial $B$-modes, this region would require stabilizing corrections to the Higgs potential in order to exit inflation without producing a patch of true vacuum.
It is notable that the lower central values for $m_t$ favored by \cite{Khachatryan:2015hba,Aaboud:2016igd} (compared to the old global value $m_t = 173.34 \GeV$ \cite{ATLAS:2014wva}) increases the amount of parameter space known to be compatible with any possible inflationary scale. However, the best-fit values would still require $H \lsim 10^{10} \GeV$ ($r \lsim 2 \times 10^{-9}$).

The limits given here are subject to some uncertainty, for instance resulting from the logarithmic running approximation employed in \Eref{eq:higgspot}. In using a quartic potential, we also neglect any quadratic terms. At large field values, the effects of the Higgs mass-squared parameter are small, but we are also assuming that terms of the form $H^2 h^2$ are suppressed as for a(n approximately) conformally coupled scalar. Note that, even in the absence of a direct Higgs coupling to curvature, Higgs couplings will radiatively generate $H^2 h^2$ terms as in \cite{Herranen:2014cua}, but the loop-sized coefficients reduce the impact of such terms (see, \eg, \Fref{fig:hsqHsq}). We return to the effect of such terms with more generic coefficients in \Sref{subsec:stabilizingcorrections}. There is also uncertainty due to the precise matching scale used for the quartic, as well as subdominant residual gauge variation as a result of the leading order approximations employed.
While the exact error is difficult to quantify, we estimate it to be $\sim 10\%$, comparable to that resulting from the exact $N_{\rm max}$ required. As such, the variation in limits between $N_{\rm max} = 50$ and $N_{\rm max} = 60$ may be taken as approximately representing the uncertainty.

\subsection{Comparison to approximate FP solutions}
\label{subsec:approximate}

For the interested reader, we now compare our results to those obtained from
employing various approximations, both in order to highlight several important
effects captured by the full FP solution, and to make connection with previous literature.

One approach, employed in several earlier works
\cite{Espinosa:2007qp,Hook:2014uia,Kearney:2015vba,Espinosa:2015qea}, is to approximate the
field distribution as Gaussian
\be
P(h,N) = \frac{1}{\sqrt{2 \pi \vev{h^2}}} \exp\left[-\frac{h^2}{2 \vev{h^2}}\right]
\ee
in the quantum-dominated regime $\abs{h} < \hcld$. For a potential $V =
\frac{\lambda h^4}{4}$ with constant $\lambda$, which we take to be
$\lambda(\mu = H)$, the variance can be computed via the equation of motion
\cite{Kearney:2015vba}
\be
\label{eq:hsqevolution}
\frac{d}{dt} \vev{h^2(t)} = \frac{2 \lambda}{H} \vev{h^2(t)}^2 + \frac{H^3}{4 \pi^2}.
\ee
The solution to this equation when $\lambda > 0$ is
\be
\label{eq:HFSoln}
\vev{h^2(t)} = \frac{1}{\sqrt{2 \lambda}} \frac{H^2}{2 \pi} \tanh\left(\sqrt{2 \lambda} \frac{N}{2 \pi}\right)
\ee
with $\tanh$ replaced by $\tan$ for $\lambda < 0$. For $\lambda \rightarrow 0$,
this reproduces the result found in \cite{Hook:2014uia}, and later in \cite{Espinosa:2015qea}, for a
negligible quartic coupling.
The probability of finding a fluctuation $\abs{h} \geq h_0$ is then given by
\begin{align}
P(h \geq h_0,N) & = 1 - \erf\left(\frac{h_0}{\sqrt{2 \vev{h^2}}}\right) \nonumber\\
& \simeq \sqrt{\frac{2 \vev{h^2}}{\pi h_0^2}} \exp\left(-\frac{h_0^2}{2 \vev{h^2}}\right).
\end{align}

Clearly, this approach does not accurately describe the behavior of large
fluctuations.  First, for $\abs{h} \gsim H$, $\lambda\left(\mu \simeq \sqrt{H^2
+ h^2}\right) \neq \lambda(\mu = H)$.  This running of $\lambda$ is subdominant
in the quantum-dominated regime $\abs{h} \lsim \hcld$, but needs to be
appropriately addressed for larger fluctuations $\abs{h} \gsim \hcld$, where
classical evolution dominates.  Second, previous implementations of this approach (such as \Rref{Hook:2014uia} and later \Rref{Espinosa:2015qea}), simply assumed that locally the field instantaneously evolves to the true vacuum once a fluctuation reaches $\abs{h} \gsim \hcld$. But, this does not appropriately account for the finite
time taken for the fluctuation to diverge.  Here, we attempt to account for this
additional time by calculating how long it takes for a patch to evolve from
$\abs{h} \simeq \hcld$ to $\abs{h} \simeq \hsrb$ under the classical equation of
motion
\be
\label{eq:classeom}
\ddot{h} + 3 H \dot{h} + V'(h) = 0,
\ee
which we denote $\Delta N_{\rm cl}$. As such, we estimate $N_{\rm max} \simeq
N_{\rm cl} + \Delta N_{\rm cl}$.\footnote{For a Gaussian distribution, this is a
reasonable approximation because the bulk of the distribution with $\abs{h}
\gsim \hcld$ is concentrated near $\abs{h} \simeq \hcld$ (larger fluctuations
being exponentially less likely). Thus, the time taken for the part of the
distribution with $\abs{h} \gsim \hcld$ to spread to $\abs{h} \gsim \hsrb$
should be approximately given by $\Delta N_{\rm cl}$.}
Typically, $\Delta N_{\rm cl} \sim 10$--$20$.

In \Fref{fig:Nelimit}, we show $N_{\rm max}$ obtained using this prescription
(left panel; blue, dotted), as well as from the similar prescription of
\Rref{Espinosa:2015qea} [Eq.~(32) therein], which uses $\lambda = 0$ and
neglects classical evolution (green, dash dotted). 
Comparing these results demonstrates the importance of both (i) the stabilizing effect of the quartic for small fluctuations (as $H < \Lmax$) and (ii) the additional time taken for a true vacuum patch to form due to the duration of the classically dominated evolution.
Together, these effects substantially extend the time taken for a true vacuum patch to form, relaxing the limit on $H$ from $H/\Lmax \lsim 0.045$ as in \cite{Espinosa:2015qea} [or $H/\Lmax \lsim 0.046$ for the central $(m_h,m_t)$ used here] to $H/\Lmax \lsim 0.065$.

However, this procedure underestimates the effect of the quartic in both regimes.
For small fluctuations in the quantum-dominated regime, the Gaussian approximation underestimates the stabilizing impact. This can be seen in the right panel of \Fref{fig:Nelimit}, which compares the full FP solution to Gaussian approximations---for the full solution, the distribution is concentrated at smaller $\abs{h}$. By itself, this would further enhance the time taken for a true vacuum patch to form; for instance, the magenta solid line of \Fref{fig:Nelimit} employs the same procedure for matching between quantum- and classical-dominated phases as for the Gaussian approximations, but uses the full FP solution for $\abs{h} \leq \hcld$. This gives the less stringent limit $H/\Lmax \lsim 0.076$.
However, that this limit is even weaker than the actual limit obtained from the full FP solution, \Eref{eq:conservativelimit}, reveals that the quartic also accelerates the growth of large fluctuations relative to the classical expectation, resulting in elongated, non-Gaussian tails of $P(h,t)$ at large fluctuations (the importance of which was first emphasized in \cite{Kearney:2015vba}). These tails are also visible in the right panel of \Fref{fig:Nelimit} and mean that, by the time we expect a patch with $\abs{h} > \hcld$ to have formed, this patch is not so overwhelmingly likely to have $\abs{h} \simeq \hcld$ as opposed to some larger value (which would take less time to diverge to the true vacuum). As such, it does not take the full $\Delta N_{\rm cl}$ $e$-folds for a true vacuum patch to form, so the actual limit is slightly more stringent than $H/\Lmax \lsim 0.076$.
Likewise, the slow falloff of the non-Gaussian tails of the distribution in the range $\abs{h} > \Lmax$, coupled with the exponentially increased volume from inflation, is the source of the similarity between Eqs.~\ref{eq:conservativelimit} and \ref{eq:nostablelimit}. 

As a final point of comparison, we note that a Hawking-Moss (HM) calculation gives the probability for the Higgs field in a Hubble patch to transition to the top of the potential barrier
\be
P_{\rm HM} \simeq \exp\left[-\frac{8 \pi^2 V(\Lmax)}{3 H^4}\right].
\ee
Requiring that no patches transition out of the EW vacuum via a HM instanton within $N_{\rm max} = 60$ $e$-folds of inflation gives the limit $H/\Lmax \lsim 0.061$, in good agreement with \Eref{eq:nostablelimit}. This provides a useful consistency check, since the FP approach should reproduce the HM transition probability in the $H \ll \Lmax$ regime \cite{Hook:2014uia}.


Overall, we find that, in the presence of a Higgs vacuum instability, the
existence of our Universe requires that any inflationary epoch satisfy $H \lsim
0.07 \Lmax$. Moreover, we note that this result is fairly insensitive to
postinflationary physics; while the constraint does weaken if we suppose
fluctuations beyond the barrier are stabilized by, \eg, efficient reheating, the
long, non-Gaussian tails of the fluctuation probability distribution make this
effect small.\,\footnote{Points at the tail of the distribution exit the slow-roll regime, and diverge rapidly to the true vacuum and backreact on spacetime  {\it{within a single Hubble time}}; it is unlikely that any restoring preheating/reheating dynamics, however extreme, can come into full effect on such short time scales.}

\subsection{Effect of stabilizing correction to the Higgs potential}
\label{subsec:stabilizingcorrections}
Finally, we comment on the possibility of additional Higgs couplings to
inflationary dynamics that may be capable of sufficiently stabilizing the Higgs potential during inflation.  For instance, Higgs-inflaton and Higgs-curvature
couplings are generally induced by loop corrections
\cite{Herranen:2014cua,Gross:2015bea} and have been suggested as a minimal
stabilization mechanism
\cite{Espinosa:2007qp,Lebedev:2012sy,Hook:2014uia,Enqvist:2014bua,Herranen:2014cua,Kamada:2014ufa,Espinosa:2015qea}
of the EW vacuum during inflation because of their contribution to the effective
mass of the Higgs. Similarly, Planck-suppressed operators coupling the Higgs to
the inflaton or the inflaton potential can result in a large effective mass
\cite{Hook:2014uia}, \eg,
\be
\label{eq:HsqfromHDO}
V \supset \frac{k V_I h^2}{M_P^2} = 3 k H^2 h^2,
\ee
which for $k > 0$ would stabilize the vacuum at $h = 0$.\footnote{A related
alternative is that Higgs couplings to moduli may modify and stabilize the
potential as in, \eg, \cite{Ema:2016ehh}. Here we focus on couplings
directly to $H$.}

Using the methods outlined above, we can determine the importance of such
additional terms in the Higgs potential for delaying the development of a patch
of true vacuum.  We will simply consider adding a term of the form
\be
V \supset \frac{c_1}{2} H^2 h^2
\ee 
to \Eref{eq:higgspot} during inflation and remain
agnostic to the source of such a term---though, as we comment below, the
underlying interaction responsible for generating this term may have important
implications. 

\begin{figure}
\begin{center}
\includegraphics[width=0.55\columnwidth,draft=false]{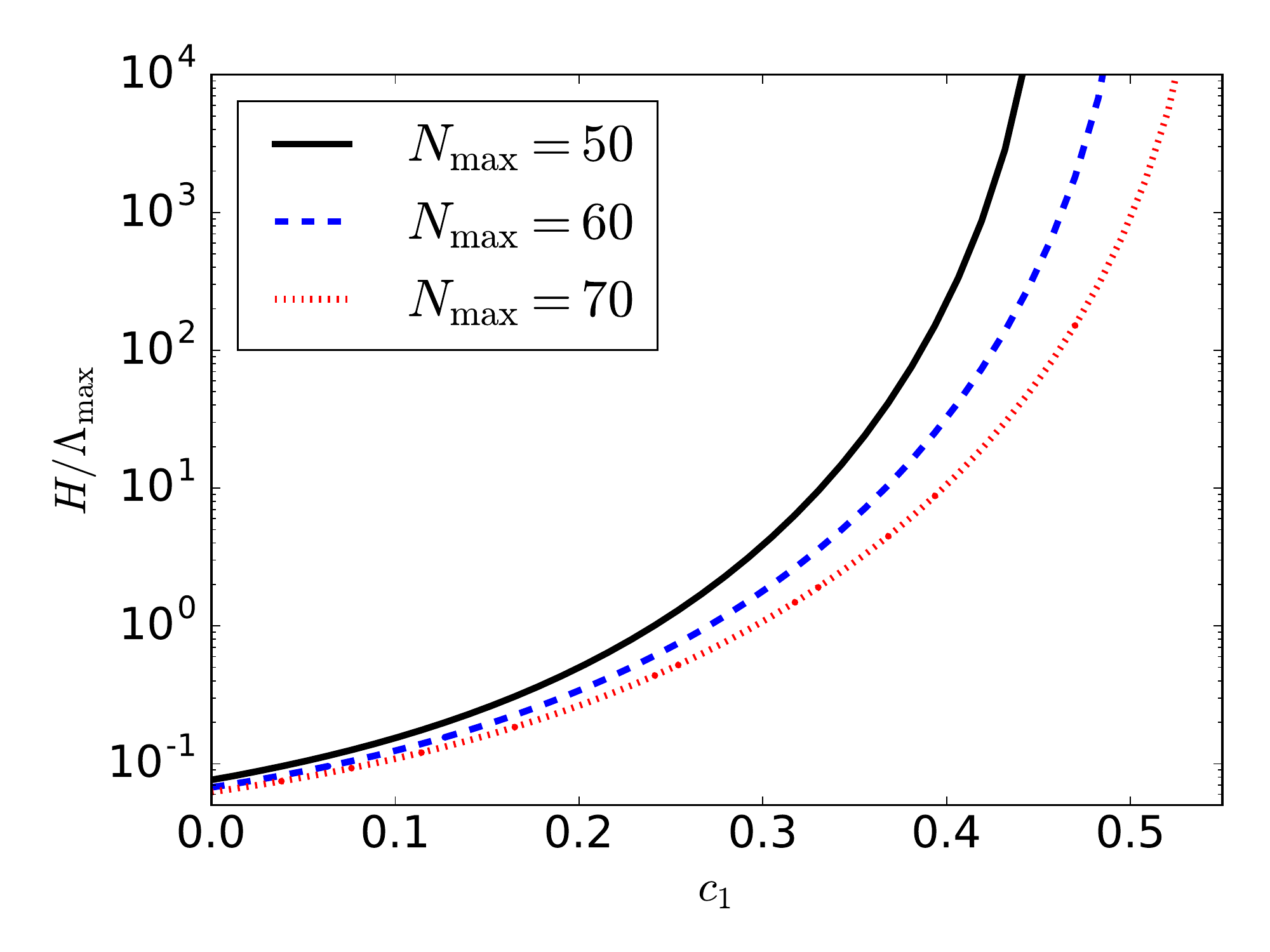}
\end{center}
\caption{
    The bounds on the ratio $H/\Lambda_{\rm max}$ when the term
    $\frac{1}{2}c_1H^2h^2$ is included in the Higgs potential, for $N_{\rm 
    max}=50$, 60, and 70 $e$-folds of inflation. 
\label{fig:hsqHsq}
}
\end{figure}

In \Fref{fig:hsqHsq}, we show how the constraint on $H/\Lambda_{\rm max}$ is
relaxed for various values of the coefficient $c_1$. For $c_1 \sim 1/4$, the
bound on the energy scale of inflation is weakened to $H \lsim \Lambda_{\rm
max}$ while, for sufficiently large values of $c_1 \gsim 1/2$, the EW vacuum becomes
effectively stable throughout inflation, such that any value of $H$ is
permissible (as anticipated from the HM calculation of \cite{Hook:2014uia}).  This is because the typical size of fluctuations goes as
$\sim \frac{H}{2 \pi} \sqrt{N}$ while the additional term stabilizes the
potential up to $h \sim \sqrt{\frac{c_1}{\abs{\lambda}}} H$.  Since the
asymptotic value of $\abs{\lambda}$ is small for the SM Higgs ($\abs{\lambda} \lsim
0.01$), even a modest coefficient $c_1$ can result in a rapidly weakening
constraint on $H/\Lambda_{\rm max}$.  Note that here, as $\mu \simeq \sqrt{H^2 +
h^2}$ varies over a number of orders of magnitude, the logarithmic
approximation employed in \Eref{eq:higgspot} is no longer valid. Therefore 
to obtain these results, we use the full running coupling. A negative
value of $c_1$ would of course have the opposite effect, destabilizing the Higgs
potential.

It is worth noting that, depending on the source of this coupling, the
coefficient $c_1$ cannot be arbitrarily large, as such couplings may destabilize
the Higgs field after inflation \cite{GarciaBellido:2008ab,Herranen:2015ima}.
Specifically, after inflation, the Universe typically undergoes a period of
preheating, during which the inflaton oscillates with large amplitude. These
oscillations can induce large Higgs fluctuations via parametric resonance through
the same coupling $c_1$ responsible for stabilizing the Higgs during
inflation. Sufficiently large fluctuations would generate a negative effective
Higgs mass and tachyonic instability, triggering EW vacuum decay, and this implies
an upper bound on $c_1$ \cite{Ema:2016kpf,Kohri:2016wof}. For instance,
supposing that the Higgs couples to the inflaton $\phi$ as $V \supset
\frac{c^2}{2} \phi^2 h^2$ and that the inflaton oscillates with chaotic
inflationlike parameters ($m_\phi \simeq 10^{13} \GeV$ and initial amplitude
$\phi_0 \simeq M_P$), the analysis of \cite{Ema:2016kpf} would constrain $c \lsim 10^{-4}$ or $c_1 \lsim {\cal O}(10^3)$.

While this is an important constraint, it is clear that, if the Higgs-inflaton
coupling were to arise from an operator like \Eref{eq:HsqfromHDO}, such large
values would require that this operator was generated with a significantly
larger coefficient than the ${\cal O}(1)$ value expected in effective field
theory, or that the cutoff was somewhat below the Planck scale.  In addition,
the full details of the preheating and reheating phases are complicated. Notably,
interactions of the Higgs field with SM particles produced via perturbative or
nonperturbative Higgs decays
\cite{Enqvist:2013kaa,Enqvist:2014tta,Figueroa:2015rqa,Enqvist:2015sua,Lozanov:2016pac}
would result in finite density (or thermal) corrections that tend to stabilize
the Higgs effective potential.  As such, efficient thermalization may relax the
bounds or even prevent EW vacuum decay during preheating.  While these effects
were estimated in \cite{Ema:2016kpf}, further dedicated numerical studies may be
required to determine the exact bounds on Higgs-curvature or Higgs-inflaton
couplings.

\section{Conclusions}
\label{conclusions}

We have studied the dynamical response of inflating spacetime to unstable
fluctuations in the Higgs field with numerical simulations of Einstein gravity.  Our results offer, for the first time, an in-depth understanding of how spacetime evolves as a Higgs fluctuation falls towards, and eventually reaches,
the true, negative energy, vacuum. We find that when true vacuum patches stop inflating and create a
crunching region, and the energy liberated creates a black hole surrounded by a shell of
negative energy density. This region of true vacuum persists and grows
throughout inflation, with more and more energy being locked behind the black
hole horizon. In contrast to the na\"ive expectation that this growth is
due to the boundary between true and metastable vacua sweeping outward in
space, in an exponentially expanding spacetime the growth occurs in a
causally disconnected manner. Spatial points fall to the true vacuum
independent of the fact that neighboring points have also reached the true
vacuum. Hence, under most circumstances, this process is insensitive to the behavior in the interior
region, and to the exact shape of the potential close to the true minimum.

We also explored nonspherically-symmetric solutions, where, in addition to
confirming that the results from the spherically symmetric case apply more generally,
we found that the formation of black holes with arbitrarily elongated horizons,
or even black strings, was possible, in violation of the hoop conjecture. As
such, the Higgs instability provides a quite different setting---one
proceeding from an initially dS-like spacetime---where some of the exotic
features seen in AdS-like spacetimes are realized.

We also extended the numerical solution of the Fokker-Planck equation to resolve
the field distribution in the exponentially suppressed tails.  This is necessary
to extract the tiny probabilities associated with a single true vacuum patch in
our past light cone, while simultaneously incorporating the effects from
renormalization group running of the quartic in the Higgs potential on the
evolution of the probability distribution. Using this solution, in conjunction
with the result from our classical General Relativity simulations that a single
true vacuum patch in our past light cone destroys the Universe, we derived a
bound $H/\Lmax \lsim 0.07$ on the scale of inflation. This bound is the most
accurate available to date, and we compared it to bounds derived previously. We
also found, as shown in \Fref{fig:HIlimit}, that a future measurement of the
tensor to scalar ratio with $r > 0.002$ would imply the need for a stabilizing
correction to the Higgs potential at a scale $\; \lsim 10^{14} \GeV$ supposing
$m_t \gsim 171.4 \GeV$.  We are thus able to correlate a cosmological quantity
with the necessity of stabilizing corrections to the Higgs potential.

Finally, we reemphasize that the results in this paper are of wider interest than the SM Higgs potential, as they are applicable to the
inflationary dynamics of any scalar field with a negative energy true vacuum.

\acknowledgments
We thank the Universe for surviving. JK also thanks Andrew Long for helpful
discussions.  Simulations were run on the Sherlock Cluster at Stanford
University.  This research was supported in part by Perimeter Institute for
Theoretical Physics. Research at Perimeter Institute is supported by the
Government of Canada through the Department of Innovation, Science and Economic
Development Canada and by the Province of Ontario through the Ministry of
Research, Innovation and Science.  JK is supported by the DoE under contract
number DE-SC0007859 and Fermilab, operated by Fermi Research Alliance, LLC
under contract number DE-AC02-07CH11359 with the United States Department of
Energy, and gratefully acknowledges the Aspen Center for Physics, which is
supported by National Science Foundation grant PHY-1066293, where part of this
work was performed. BS is supported by the DoE under grants DE-SC0007859 and
DE-SC0011719.  KZ is supported by the DoE under contract DE-AC02-05CH11231.

\appendix

\section*{Appendix: Details of Numerical Methods}
\label{app:numericaldetails}
When solving the Einstein equations, for the metric initial data we use a
conformally flat spatial metric $\gamma_{ij}=\Psi^4 f_{ij}$ and set the trace of
the extrinsic curvature according to the inflationary Hubble parameter
$K=-3H$, while fixing the traceless part to be zero.  With these choices, the
momentum constraint is trivially satisfied, while the Hamiltonian constraint is
solved using the code described in~\cite{idsolve_paper} to obtain the conformal
factor $\Psi$.  In practice, since we consider cases with $h_{\rm in}\ll h_{\rm
min}$, the conformal factor is always close to unity, and hence the initial
metric is very nearly just a slice of de Sitter in planar coordinates.

We evolve the Einstein field equations in the generalized harmonic formulation
as described in~\cite{Pretorius:2004jg, code_paper}.  In this formulation, the
coordinate degrees of freedom are specified through the source functions
$\Box x_a = H_a$.  Here we fix the source functions to be those of the
inflationary de Sitter metric: $H_t=3H$ and $H_i=0$.  We use compactified
coordinates that extend to spatial infinity where we impose the boundary
condition that the metric be exactly de Sitter. Hence, away from any regions
with large Higgs field fluctuations or potential energy, the coordinates
$\{t,x^i\}$ we use will very closely match de Sitter planar coordinates.

\begin{figure}
\begin{center}
\hskip-0.2in
\includegraphics[width=0.5\columnwidth,draft=false]{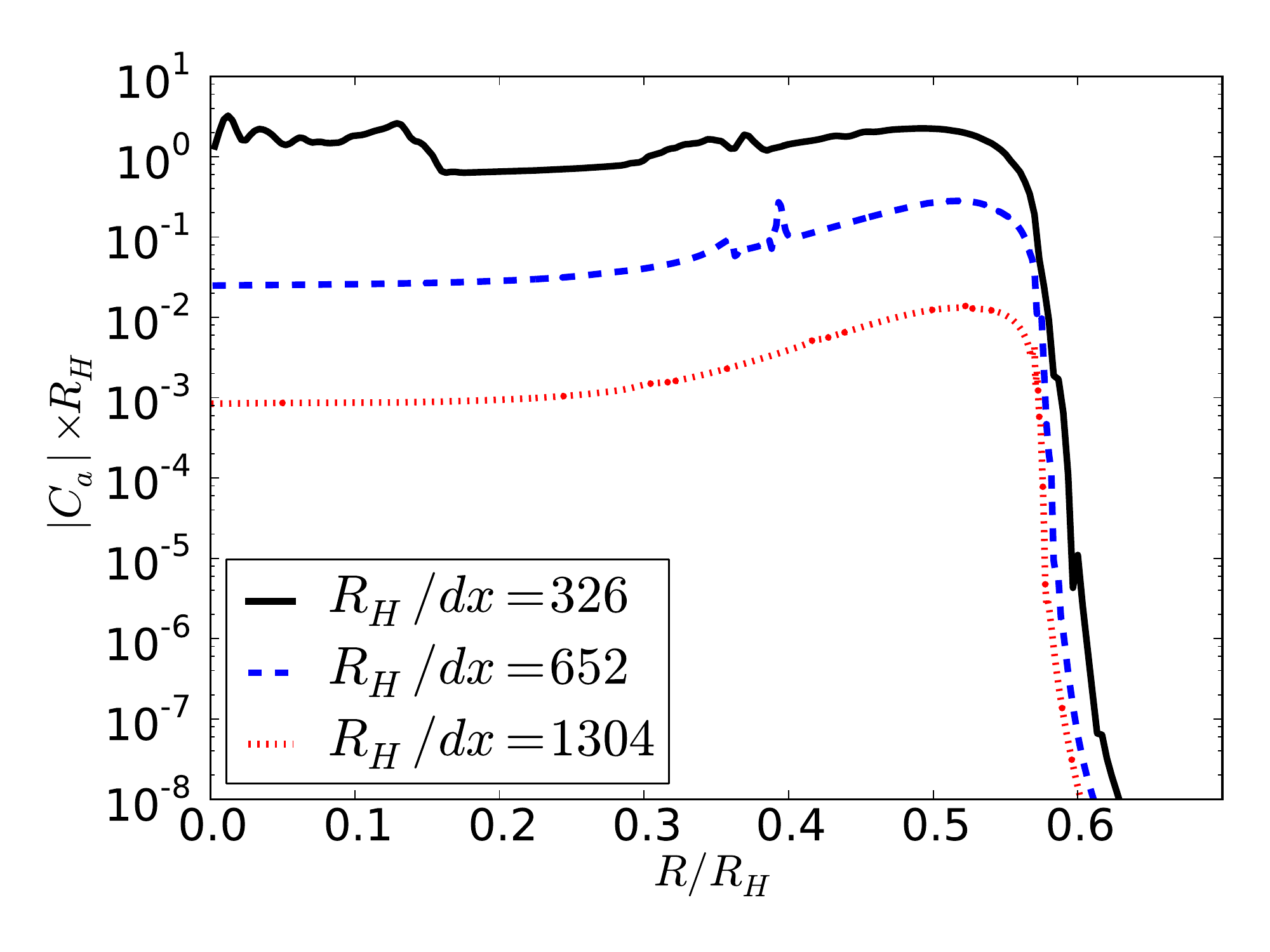}
\includegraphics[width=0.5\columnwidth,draft=false]{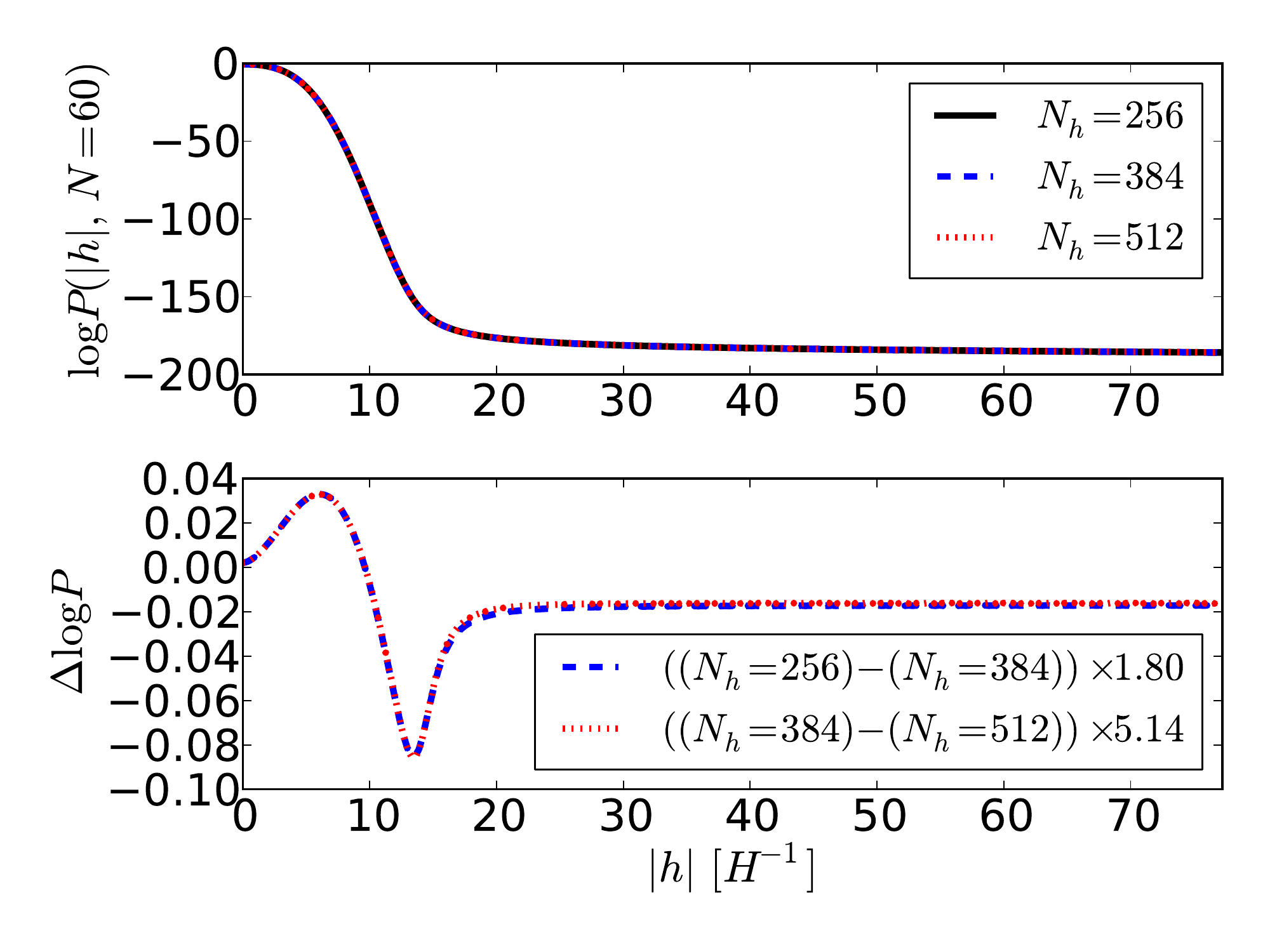}
\end{center}
\caption{ 
Left panel: The norm of the Einstein equation constraint violation $C_a=\Box
x_a -H_a$ for simulations at three different resolutions of the case shown in
\Fref{fig:bubble}, at $tH=0.78$ (just before the formation of a black hole).
The scaling with resolution is consistent with approximately fourth-order
convergence.  
Right panel: Solutions of the Fokker-Planck equation [\Eref{eqn:FPX}] for
$H/\Lmax=0.07$ and $b_0=0.16/(4\pi)^2$ at $N=60$ at three different resolutions
(top), and the difference between the resolutions (bottom).  In the latter case,
the quantities are scaled to indicate the error in $\log P$ for the lowest
resolution (which is used for the results in this paper), consistent with
second-order convergence.   
\label{fig:conv} }
\end{figure}

As in~\cite{East:2015ggf}, we evolve both the metric and scalar field using
fourth-order finite differences and fourth-order Runge-Kutta time stepping.  We
take advantage of the axisymmetry of the problem to make the computational grid
a half-plane while still using Cartesian coordinates through the use of a
modified Cartoon method as described in~\cite{Pretorius:2004jg}.  To eliminate
numerical error coming from just evolving the known de Sitter solution, we use
the background error subtraction technique~\cite{East:2013iwa}.  In the left panel of
\Fref{fig:conv}, we demonstrate the expected convergence for an example
case.  The results presented in this paper use the medium or high resolution
shown there.

We solve the FP equation written in terms of the variable $X \equiv \log P$,
\Eref{eqn:FPX}, using standard second-order finite differences for the field
derivatives, and the Crank-Nicholson method for the time integration---an implicit method often used for diffusion type equations.  
As an initial condition, we choose $P$ to be a narrow Gaussian distribution with zero
mean and variance given by \Eref{eq:HFSoln} evaluated at $N=1/8$.  At the
outer boundary, we impose the condition that $\partial X / \partial
h=h(\partial^2 X / \partial^2 h)$, which is chosen to be compatible with a
Gaussian initial condition.  We have verified that our results are not sensitive
to the placement of the outer boundary (at a few times $\hsrb$) or the exact
width of the initial distribution. The numerical error and convergence for 
an example case are shown in the right panel of \Fref{fig:conv}.

\bibliographystyle{h-physrev}
\bibliography{EOH_refs}

\end{document}